\documentclass[12pt]{elsarticle}

\usepackage{amssymb}

\usepackage{graphicx}
\usepackage{amsmath}
\usepackage{subfigure}
\usepackage{color}
\usepackage{multirow}
\biboptions{sort&compress}

\journal{Journal of magnetism and magnetic materials}

\begin{document}

\begin{frontmatter}



\title{Large-$s$ expansions for the low-energy parameters of the
    honeycomb-lattice Heisenberg antiferromagnet with spin quantum
    number $s$}


\author{R.F. Bishop and P.H.Y. Li}
\ead{raymond.bishop@manchester.ac.uk; peggyhyli@gmail.com}

\address{School of Physics and Astronomy, Schuster Building, The University of Manchester, Manchester, M13 9PL, UK}

\begin{abstract}
The coupled cluster method (CCM) is employed to very high orders of approximation to study the ground-state (GS) properties of the spin-$s$ Heisenberg antiferromagnet (with isotropic interactions, all of equal strength, between nearest-neighbour pairs only) on the honeycomb lattice.  We calculate with high accuracy the complete set of GS parameters that fully describes the low-energy behaviour of the system, in terms of an effective magnon field theory, viz., the energy per spin, the magnetic order parameter (i.e., the sublattie magnetization), the spin stiffness and the zero-field (uniform, transverse) magnetic susceptibility, for all values of the spin quantum number $s$ in the range $\frac{1}{2} \leq s \leq \frac{9}{2}$.  The CCM data points are used to calculate the leading quantum corrections to the classical ($s \rightarrow \infty$) values of these low-energy parameters, considered as large-$s$ asymptotic expansions. \end{abstract}


\end{frontmatter}


\section{INTRODUCTION}
\label{introd_sec}
Isotropic Heisenberg antiferromagnets (HAFs) have a continuous SU(2)
rotational symmetry in spin space, which may be spontaneously broken
under rather general conditions down to its U(1) subgroup, thereby
leading to classical ground-state (GS) phases with magnetic long-range
order (LRO).  Such states are typically not eigenstates of the
corresponding quantum Hamiltonian in the case where the spins have a
finite value of the spin quantum number $s$.  The role then played by
quantum fluctuations on the corresponding GS ordering properties of
such HAFs comprising interacting quantum spins placed on the sites of
an (infinite) regular periodic lattice continues to engender
considerable interest, both theoretically and experimentally.

In very general terms, quantum fluctuations are larger for systems
with lower dimensionality $D$, lower values of the spin quantum number
$s$, and lower values of the coordination number $z$ of the spatial
lattice.  The Mermin-Wagner theorem \cite{Mermin:1966}, which asserts
the impossibility of breaking a continuous symmetry when $D=1$, even
for systems at zero temperature ($T=0$), thus precludes GS phases with
magnetic LRO for one-dimensional (1D) spin chains.  The same theorem
also rules out magnetic LRO in any isotropic system with $D=2$ at any
nonzero temperature ($T > 0$).  Since it does not, however, apply to
2D systems at $T=0$ (or, indeed, to systems with $D > 2$), 2D quantum
magnets at $T=0$ provide a key arena for the study of the role of
quantum fluctuations on their properties.  Furthermore, since the
honeycomb lattice has the lowest coordination number ($z=3$) of all
regular 2D lattices, it is natural to focus particular attention on
it, as we do here.

The behaviour at low energies or large distances of any strongly
correlated system that has undergone spontaneous symmetry breaking is
governed by the properties and dynamics of the massless Goldstone
bosons that thereby emerge \cite{Goldstone:1961_boson}.  In the case of the isotropic HAFs
considered here these are simply the spin waves or magnons.  In turn,
the dynamics of the Goldstone bosons can be precisely formulated in
terms of a simple, systematic effective field theory (EFT)
\cite{Chakravarty:1989_magnon-chiral-PT,Neuberger:1989_magnon-chiral-PT,Fisher:1989_magnon-chiral-PT,Hasenfratz:1990_magnon-chiral-PT,Hasenfratz:1991_magnon-chiral-PT,Hasenfratz:1993_magnon-chiral-PT,Chubukov:1994_magnon-chiral-PT},
which is specified wholly by the symmetry properties of the model in
terms of a few low-energy parameters.  While models in the same
symmetry class are thus described by a universal EFT, the values of
the low-energy parameters themselves depend on the specific model
being studied.  Thus, while a particular EFT (pertaining to a given
symmetry class) leads to universal expressions for such asymptotic
formulae or scaling forms as those pertaining to finite-size or
low-temperature corrections, one still needs an independent knowledge
of the values of the low-energy parameters to
implement them in practice.

In principle the parameters can be obtained in two different ways.  In
the first one performs a microscopic calculation at finite values of
the system size and/or at nonzero temperatures ($T=0$), and uses the
results as input to the EFT scaling forms to extract the low-energy
parameters.  In the second way one simply calculates the low-energy
parameters directly for a system of infinite size (i.e., in the
appropriate thermodynamic limit) and at $T=0$, using some suitable
{\it ab initio} technique of microscopic quantum many-body theory.
For the specific case of the spin-$\frac{1}{2}$ HAF, with
neareast-neighbour (NN) interactions only, on the honeycomb lattice,
typical calculations of the former type have been performed using
quantum Monte Carlo (QMC) algorithms of various types
\cite{Castra:2006_honey,Low:2009_honey,Jiang:2012_honey}.  Although
QMC calculations can be highly accurate they are often restricted in
practice to unfrustrated systems, i.e., in the present case to models
with NN interactions only, due to the well-known ``minus-sign
problem''.  An alternative technique of the former sort, which uses
the exact diagonalization (ED) of finite lattices, does not suffer
from the same restriction, but is restricted in practice to much
smaller systems, which are hence also more problematic in fitting to
the asymptotic finite-size scaling forms.

There are relatively few microscopic spin-lattice techniques of the
second sort that can be applied to systems of $N$ spins from the
outset in the limit $N \rightarrow \infty$.  Among them are the
linked-cluster series expansion (SE) method
\cite{Rigo:2006_honey,Oitmaa:2006_honey} and the coupled cluster
method (CCM)
\cite{Bishop:1987_ccm,Bartlett:1989_ccm,Arponen:1991_ccm,Bishop:1991_TheorChimActa_QMBT,Bishop:1998_QMBT_coll,Zeng:1998_SqLatt_TrianLatt,Fa:2004_QM-coll}.
Both methods have been applied, for example, to calculate the
low-energy parameters of the spin-$\frac{1}{2}$ HAF on the honeycomb
lattice (see, e.g., Ref.\ \cite{Oitmaa:1992_honey} for an SE
calculation and Ref.\ \cite{Bishop:2015_honey_low-E-param} for a CCM
calculation that also includes frustrating bonds).

Another technique that is commonly applied to spin-lattice problems,
and which is complementary to those discussed above is spin-wave
theory (SWT) \cite{Anderson:SWT,Kubo:1952_SqLatt,Oguchi:1960_SqLatt}.
It essentially works best close to the classical limit ($s \rightarrow
\infty$), and develops series expansions in powers of $1/s$ for the
low-energy parameters.  In this context one of the strengths of the CCM
in particular is that it is relatively straightforward both in
principle and in practice to apply to models with arbitrary values of
the spin quantum number, $s$.  One of the main purposes of the present
paper is thus to apply the CCM to high orders of approximation to
study the GS properties of the honeycomb-lattice HAF with values of
the spin quantum number in the range $1 \leq s \leq \frac{9}{2}$, with
a particular aim to examine the asymptotic large-$s$ expansions for
the low-energy parameters that describe the model via EFT.  We note that there is no fundamental reason to limit our calculations to the cases with $s \leq \frac{9}{2}$.  The CCM can readily also be applied to spin-lattice models with much higher values of $s$.  The choice to limit ourselves here to cases with $s \leq \frac{9}{2}$ is made purely on the practical ground that this range surely suffices both to highlight the efficacy of the CCM and to investigate fully the evolution of the low-energy parameters as a function of increasing spin quantum number $s$, which are our joint main aims.

The outline of the rest of the paper is as follows.  We first describe
in Sec.\ \ref{low-E-param_sec} the low-energy parameters that describe
the magnon EFT.  The CCM technology that we use to calculate them, and
the hierarchical approximation scheme that we employ, are then
outlined in Sec.\ \ref{ccm_sec}.  The method is applied to the
spin-$s$ honeycomb-lattice HAF for values of the spin quantum number
$s \leq \frac{9}{2}$, and we cite extrapolated results in Sec.\
\ref{results_sec} for the corresponding low-energy parameter set in
each case.  We use these sets to derive respective expansions in
powers of $1/s$ for each parameter about the corresponding classical
($s \rightarrow \infty$) limit and, where possible, we compare with
results from SWT.  Finally, we summarise in Sec.\ \ref{summary_sec}.

\section{LOW-ENERGY PARAMETERS}
\label{low-E-param_sec}
The systematic low-energy EFT for magnons
\cite{Chakravarty:1989_magnon-chiral-PT,Neuberger:1989_magnon-chiral-PT,Fisher:1989_magnon-chiral-PT,Hasenfratz:1990_magnon-chiral-PT,Hasenfratz:1991_magnon-chiral-PT,Hasenfratz:1993_magnon-chiral-PT,Chubukov:1994_magnon-chiral-PT}
was itself developed soon after the introduction of, and in complete
analogy to, chiral perturbation theory ($\chi$PT) (see, e.g., Ref.\
\cite{Leutwyler:1994_chiral-PT} and references cited therein) for the
pions that play the role of the Goldstone bosons in quantum
chromodynamics (QCD).  Just as the hadronic vacuum plays the role of
the broken phase in QCD, so does the antiferromagnetic (AFM) phase
play the same role for a HAF.  Similarly, just as the order parameter
is given by the chiral condensate in QCD, for a HAF it is given by the
average local on-site magnetization (or, here, equivalently the
staggered magnetization), $M$.  In the case of QCD the overall
coupling strength in $\chi$PT is given by the pion decay constant,
whereas for a HAF it is given in its EFT by the spin stiffness (or
helicity modulus), $\rho_{s}$.  Finally, whereas in QCD the
propagation speed inherent in $\chi$PT is the speed of light, for a
HAF the magnons of its EFT propagate with the corresponding spin-wave
velocity, $c$.  The latter two quantities for a HAF are related by the
effective description of spin waves by a hydrodynamic theory
\cite{Halperin:1969_SWT,Cherryshev:2009_SWT}, which yields
\begin{equation}
\hbar c = \sqrt{\frac{\rho_{s}}{\chi}}\,, \label{eq_hbar_c}
\end{equation}
where $\chi$ is the zero-field (uniform, transverse) magnetic
susceptibility, in units where the gyromagnetic ratio $g\mu_{B}/\hbar
= 1$.

Thus, the fundamental low-energy parameter set that describes
completely the low-energy physics of a magnetic system of the AFM type
considered here consists of: (a) the GS energy per particle, $E/N$,
(b) the average local on-site magnetization, $M$, (c) the zero-field,
uniform, transverse magnetic susceptibility, $\chi$, (d) the spin
stiffness, $\rho_{s}$, and (e) the spin-wave velocity, $c$.  The
latter three quantities are related via the hydrodynamic relation of
Eq.\ (\ref{eq_hbar_c}).  We note too that the parameters $\rho_{s}$
and $\chi$, in particular, are defined here per unit site, as is usual
for a discrete lattice description.  By contrast, in a continuous EFT
description, it is more normal to define corresponding quantities,
$\bar{\rho}_{s}$ and $\bar{\chi}$, per unit area.  If we define the NN
spacing on the honeycomb lattice to be $d$, the lattice then has
$4/(3\sqrt{3}d^{2})$ sites per unit area, and hence
\begin{equation}
\rho_{s}=\frac{3}{4}\sqrt{3}d^{2}\bar{\rho}_{s}\,, \quad \chi=\frac{3}{4}\sqrt{3}d^{2}\bar{\chi}\,.  \label{eq_rho_chi}
\end{equation}

\begin{figure}[t]
\begin{center}
\mbox{
\subfigure[]{\includegraphics[width=3.5cm]{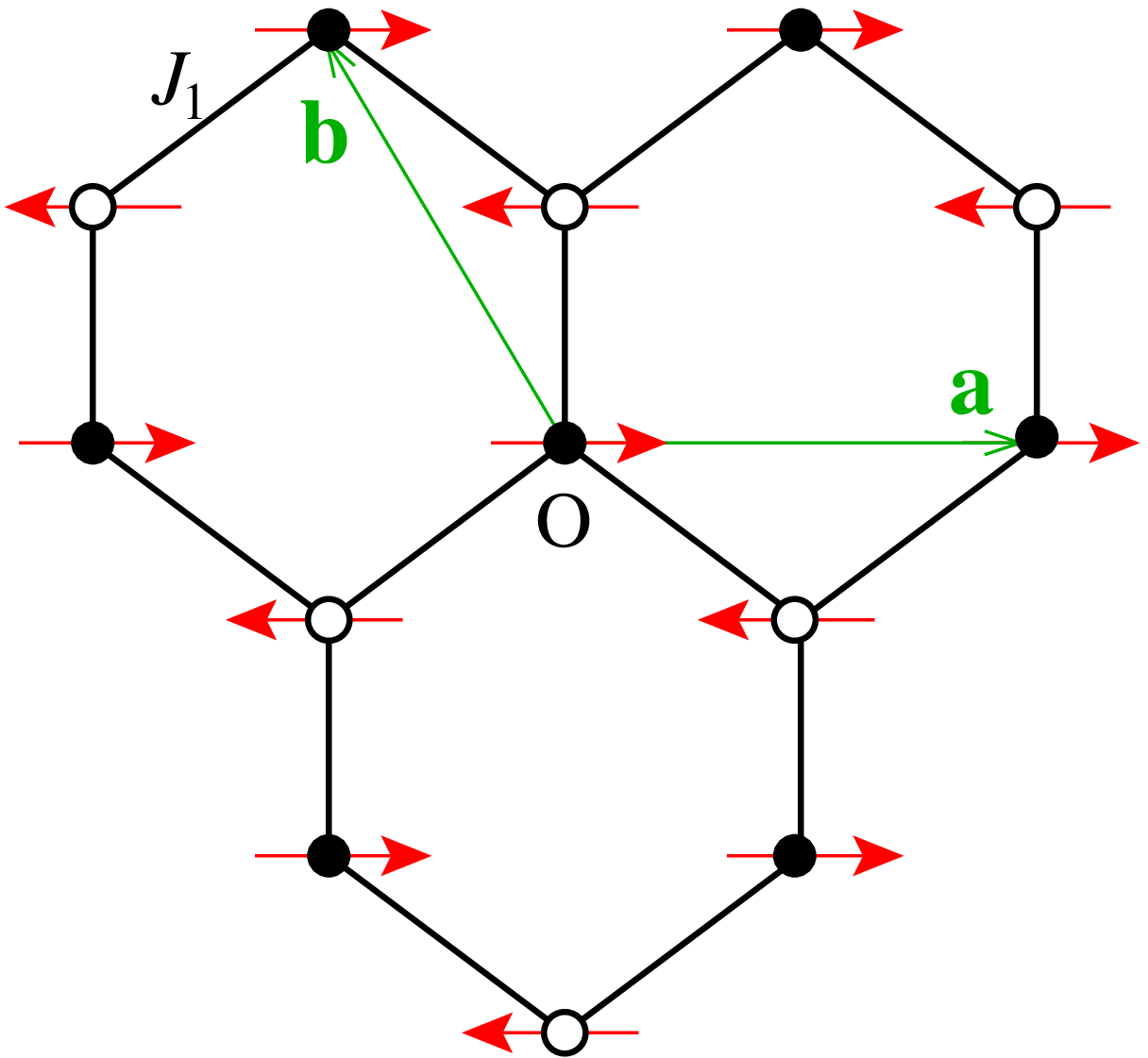}}
 \subfigure[]{\includegraphics[width=3.5cm]{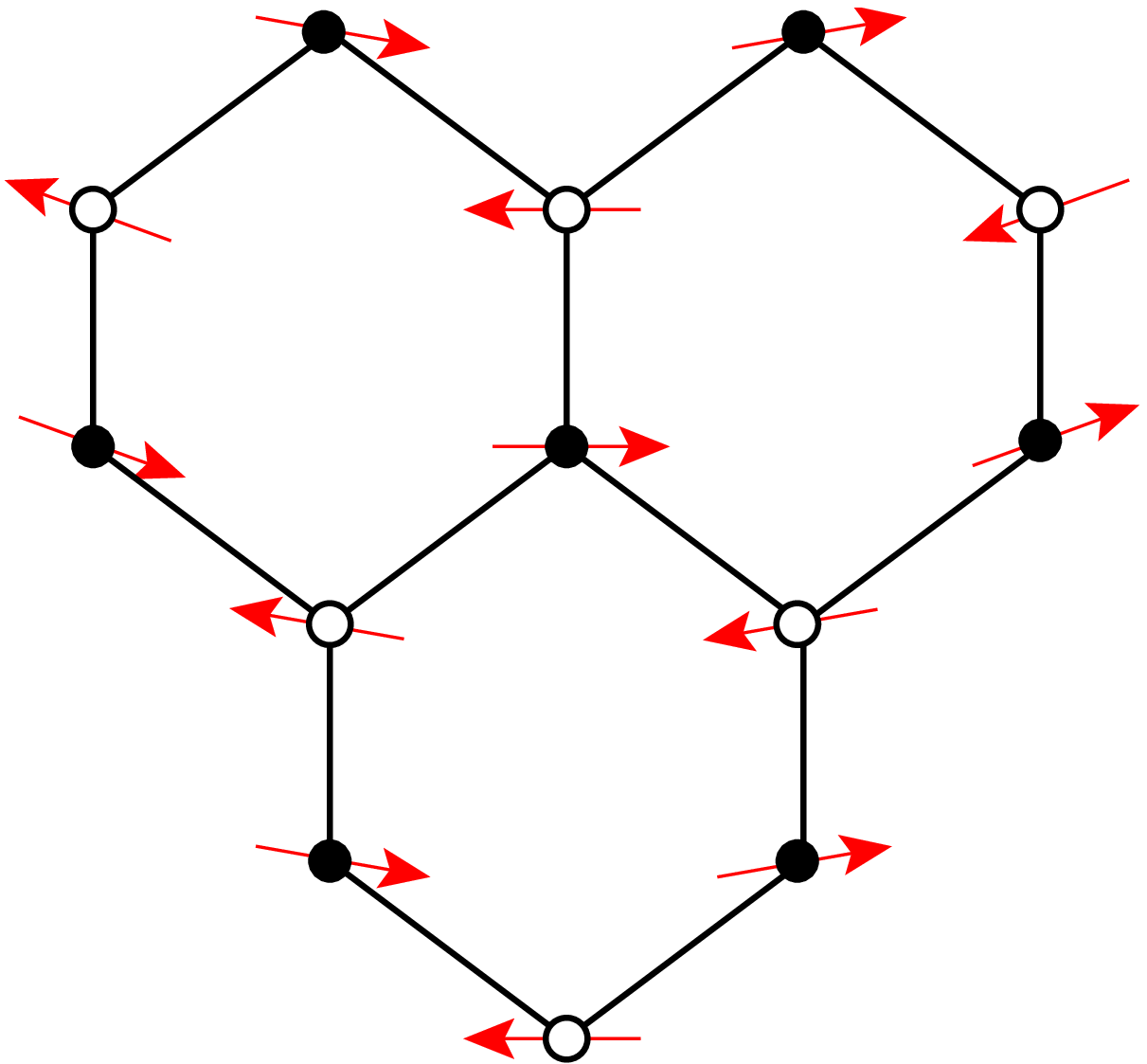}}
 \subfigure[]{\includegraphics[width=3.5cm]{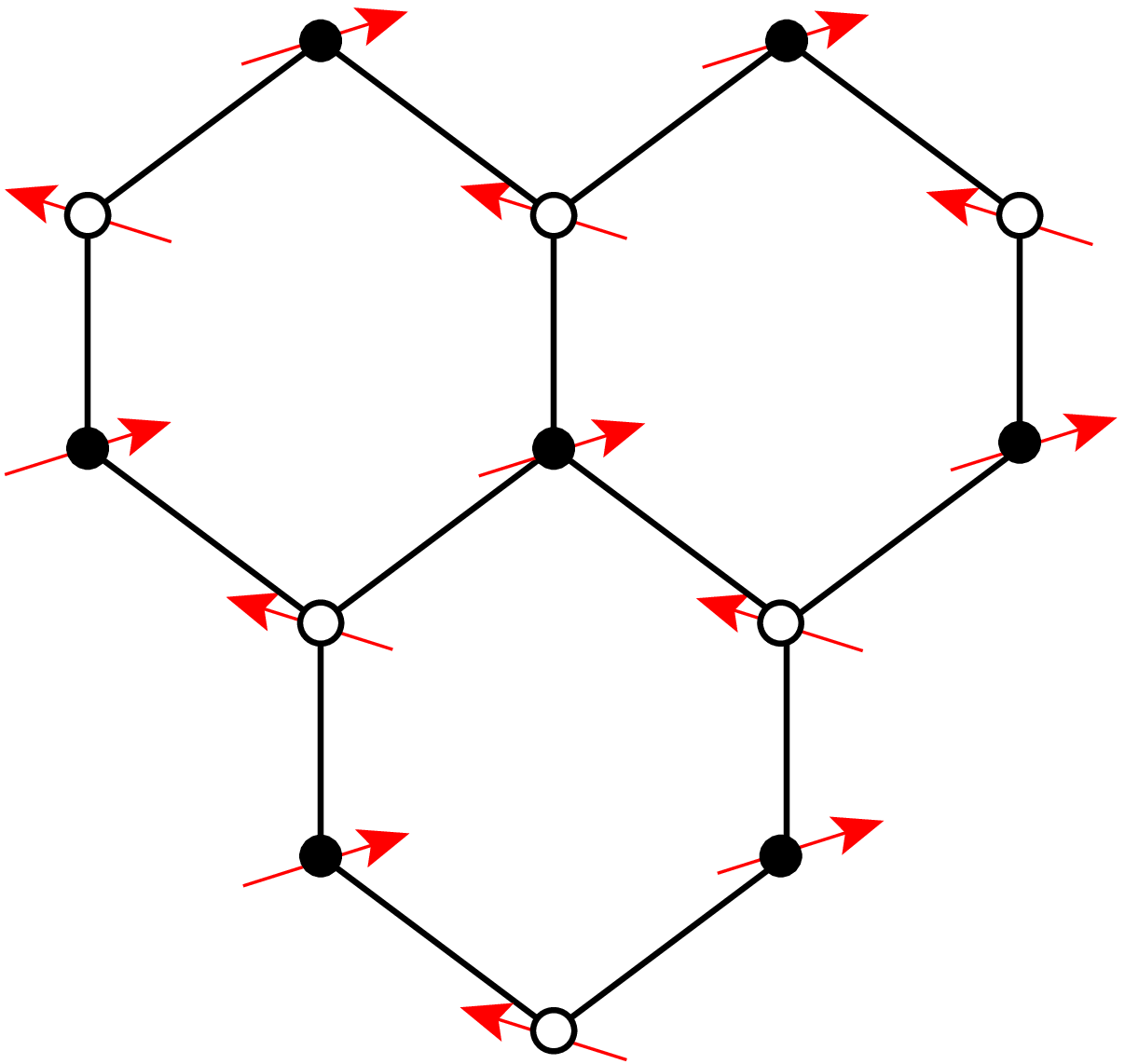}}
 \subfigure{\includegraphics[width=2.5cm]{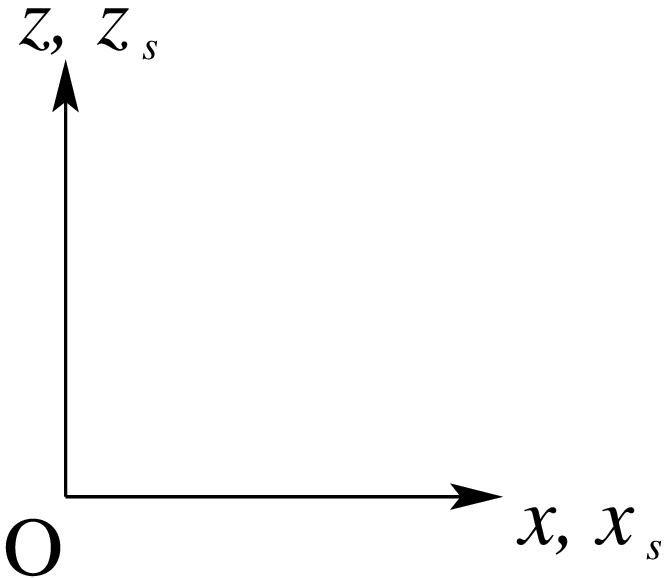}}
  }
  \caption{The HAF on the honeycomb
    lattice, showing (a) the bonds ($J_{1} = $ -----), the
    triangular Bravais lattice vectors $\mathbf{a}$ and $\mathbf{b}$,
    and the N\'{e}el state, (b) the twisted N\'{e}el state for the calculation of the spin stiffness coefficient, $\rho_{s}$, showing the twist applied in the $x$ direction, (c) the canted N\'{e}el state for the calculation of the zero-field magnetic susceptibility, $\chi$, with the external magnetic field applied in the $z_{s}$ direction.  Sites on the two
    triangular sublattices ${\cal A}$ and ${\cal B}$ are shown by
    filled and empty circles respectively, and the spins are
    represented by the (red) arrows on the lattice sites.}
\label{model_pattern}
\end{center}
\end{figure}

We place quantum spins $\mathbf{s}_{k} \equiv (s^{x}_{k},\,s^{y}_{k},\,s^{z}_{k})$ on the sites $k$ of a honeycomb lattice.  They obey the usual SU(2) commutation relations,
\begin{equation}
[s^{a}_{k},s^{b}_{l}] = i\delta_{kl}\epsilon_{abc}s^{c}_{k}\,,  \label{SU2_comm_relatn}
\end{equation}
with $\mathbf{s}^{2}_{k} = s(s+1)$ and, for the cases considered here, $s=1,\,\frac{3}{2},\,2,\,\frac{5}{2},\,3,\,\frac{7}{2},\,4,\frac{9}{2}$.  The SU(2)-invariant Hamiltonian of the quantum HAF is
\begin{equation}
H = J_{1}\sum_{\langle k,l \rangle}\mathbf{s}_{k}\cdot\mathbf{s}_{l}\,; \quad J_{1}>0\,, \label{eq_H}
\end{equation}
where the sum over $\langle k,l \rangle$ runs over all NN pairs on the
honeycomb lattice, counting each pair once only.  The Hamiltonian
commutes with the total spin operator,
\begin{equation}
[H,\mathbf{S}] = 0\,; \quad \mathbf{S} \equiv \sum^{N}_{k=1}\mathbf{s}_{k}\,.
\end{equation}
The lattice and the Heisenberg exchange bonds are illustrated in Fig.\
\ref{model_pattern}(a).

The honeycomb lattice is bipartite but non-Bravais.  It comprises two
triangular Bravais sublattices ${\cal A}$ and ${\cal B}$.  Sites on
sublattice ${\cal A}$ are at positions
$\mathbf{R}_{k}=m\mathbf{a}+n\mathbf{b}=\sqrt{3}(m-\frac{1}{2}n)d\hat{x}+\frac{3}{2}nd\hat{z}$,
where $m,n\,\in\mathbb{Z}$, in terms of Bravais lattice vectors
$\mathbf{a}\equiv\sqrt{3}d\hat{x}$ and
$\mathbf{b}=\frac{1}{2}d(-\sqrt{3}\hat{x}+3\hat{z})$, defined to lie
in the $xz$ plane, as shown in Fig.\ \ref{model_pattern}.  Each unit
cell $k$ at position vector $\mathbf{R}_{k}$ thus comprises two spins,
one at $\mathbf{R}_{k} \in {\cal A}$ and the other at
$(\mathbf{R}_{k}+d\hat{z}) \in {\cal B}$.  The honeycomb-lattice
Wigner-Seitz unit cell is thus the parallelogram formed by the lattice
vectors $\mathbf{a}$ and $\mathbf{b}$.  It may also be equivalently
taken as being centred on a point of sixfold symmetry, so that it is
bounded by the sides of a primitive hexagon of side $d$.

The appearance of a GS phase with a non-vanishing value of the
staggered magnetization order parameter,
\begin{equation}
\mathbf{M} = \frac{1}{N}\sum^{N}_{k=1}{\phi}_{k}\mathbf{s}_{k}\,,  \label{stag_M}
\end{equation}
where $\phi_{k} \equiv +1\;(-1)$ for $k \in {\cal A}\;({\cal B})$, then
signals the spontaneous breaking of the SU(2) symmetry down to its
U(1) subgroup.  The magnon field in SU(2)/U(1) may then be taken as
\begin{equation}
\mathbf{e}(\mathbf{R})=(e_{1}(\mathbf{R}),\,e_{2}(\mathbf{R}),\,e_{3}(\mathbf{R}))\,; \quad |\mathbf{e}(\mathbf{R})|^{2}=1\,; \quad \mathbf{R}\equiv(x,z)\,.
\end{equation}
The effective action for the low-energy EFT of the AFM magnons is given
by
\begin{equation}
{\cal S}[\mathbf{e}]=\int^{\beta}_{0}{\rm d}t\int{\rm d}^{2}R\;\frac{1}{2}\bar{\rho}_{s}\left(\frac{\mathbf{\partial e}}{\partial x}\cdot\frac{\mathbf{\partial e}}{\partial x}+\frac{\mathbf{\partial e}}{\partial z}\cdot\frac{\mathbf{\partial e}}{\partial z}+\frac{1}{c^{2}}\frac{\mathbf{\partial e}}{\partial t}\cdot\frac{\mathbf{\partial e}}{\partial t}\right)\,, \label{eq_low-E_EFT}
\end{equation}
in terms of the inverse temperature parameter,
$\beta\equiv\hbar/(k_{B}T)$.  We shall be interested here only in the
case $T=0$.  Note that the pre-factor of the last (temporal) term in
Eq.\ (\ref{eq_low-E_EFT}) may equivalently be written as
$\frac{1}{2}\bar{\rho}_{s}/c^{2}=\frac{1}{2}\hbar^{2}\bar{\chi}$, from
Eqs.\ (\ref{eq_hbar_c}) and (\ref{eq_rho_chi}).

The spin stiffness (or helicity modulus), $\rho_{s}$, of a
spin-lattice system is simply a measure of the energy required to
rotate the order parameter $\mathbf{M}$ of a magnetically ordered
state by an (infinitesimal) angle $\theta$ per unit length in a
specified direction.  Hence, if $E(\theta)$ is the GS energy as a
function of the imposed twist, and $N$ is the number of lattice sites,
we have
\begin{equation}
\frac{E(\theta)}{N}=\frac{E(0)}{N} + \frac{1}{2}\rho_{s}\theta^{2} + O(\theta^{4})\,.  \label{eq_GS-E_theta}
\end{equation}
We note that $\theta$ has the dimensions of an inverse length.  In the
thermodynamic limit of an infinite lattice ($N \rightarrow \infty$) a
nonzero (positive) value of $\rho_{s}$ implies the stability of the
magnetic long-range order (LRO).  For the N\'{e}el AFM state
illustrated in Fig.\ \ref{model_pattern}(a) for a staggered
magnetization in the $x_{s}$ direction, the value of $\rho_{s}$ is
completely independent of the applied twist direction.  Figure
\ref{model_pattern}(b) illustrates the twist applied in the $x$
direction to the N\'{e}el state of Fig.\ \ref{model_pattern}(a).  A
trivial calculation, using the definition of Eq.\
(\ref{eq_GS-E_theta}), shows that the value of $\rho_{s}$ for the
classical ($s \rightarrow \infty$) N\'{e}el state is
\begin{equation}
\rho^{{\rm cl}}_{s}=\frac{3}{4}J_{1}d^{2}s^{2}\,. \label{sStiff_neel_classical}
\end{equation}

Suppose we now place the N\'{e}el state shown in Fig.\
\ref{model_pattern}(a), ordered in the $x_{s}$ direction, in a
transverse uniform magnetic field, $\mathbf{h}=h\hat{z_{s}}$.  In
units where the gyromagnetic ratio $g\mu_{B}/\hbar=1$, the Hamiltonian
$H=H(h=0)$ of Eq.\ (\ref{eq_H}), then becomes
\begin{equation}
H(h)=H(0) + h\sum^{N}_{k=1} s^{z}_{k}\,.  \label{eq_H-h}
\end{equation}
The spins now cant at an angle $\alpha$ to the $x_{s}$ axis with
respect to their zero-field configurations, as shown in Fig.\
\ref{model_pattern}(c).  The classical ($s \rightarrow \infty$) value
of $\alpha$ is easily calculated by minimizing the classical energy,
$E = E(h)$, corresponding to Eq.\ (\ref{eq_H-h}), with respect to the
cant angle $\alpha$.  The uniform (transverse) magnetic susceptibility
is then defined, as usual, by
\begin{equation}
\chi({h})=-\frac{1}{N}\frac{{\rm d}^{2}E}{{\rm d}h^{2}}\,,
\end{equation}
Its zero-field limit is then the corresponding low-energy parameter,
$\chi \equiv \chi(0)$.  A simple calculation shows that the value of
$\chi$ for the classical ($s \rightarrow \infty$) N\'{e}el state is
\begin{equation}
\chi^{{\rm cl}}=\frac{1}{6J_{1}}\,. \label{chi_neel_classical}
\end{equation}
Equations (\ref{sStiff_neel_classical}) and (\ref{chi_neel_classical})
yield the corresponding classical ($s \rightarrow \infty$) limit of
the spin-wave velocity,
\begin{equation}
\hbar c^{{\rm cl}} = \frac{3}{2}\sqrt{2}J_{1}ds\,, 
\end{equation}
from Eq.\ (\ref{eq_hbar_c}), which is simply the result of lowest-order SWT (LSWT).  

\section{THE COUPLED CLUSTER METHOD}
\label{ccm_sec}
We now outline the key features of the CCM, and refer the reader to
the extensive literature (and see, e.g., Refs.\
\cite{Bishop:1987_ccm,Bartlett:1989_ccm,Arponen:1991_ccm,Bishop:1991_TheorChimActa_QMBT,Bishop:1998_QMBT_coll,Zeng:1998_SqLatt_TrianLatt,Fa:2004_QM-coll,Bishop:1978_ccm,Bishop:1982_ccm,Arponen:1983_ccm}
and references cited therein) for further details.  While the CCM was originally invented to discuss stationary states, and hence the static properties, of quantum many-body systems, it has since been extended to a fully dynamic (bi-variational) formulation \cite{Arponen:1983_ccm}, which is readily capable, both in principle and in practice, of calculating dynamic properties.  Since, however, we are only interested here in GS properites, we henceforth concentrate only on the stationary version of the formalism.  As a first step
one needs to choose a suitable many-body (normalized) model (or
reference) state $|\Phi\rangle$, in terms of which the correlations
present in the exact GS wave function $|\Psi\rangle$ can later be
systematically incorporated, in a fashion we describe below.  Although
we will describe the properties required of $|\Phi\rangle$ in detail
below, we remark now that it plays the role of a generalized vacuum
state.  For our present study the quasiclassical N\'{e}el state will
be our choice for $|\Phi\rangle$.

The exact GS ket- and bra-state wave functions, $|\Psi\rangle$ and
$\langle\tilde{\Psi}|$, respectively, are chosen to satisfy the
normalization conditions
\begin{equation}
\langle\tilde{\Psi}|\Psi\rangle = \langle{\Phi}|\Psi\rangle =
\langle{\Phi}|\Phi\rangle \equiv 1\,.   \label{norm_conditions}
\end{equation}
They are now parametrized with respect to the chosen reference state
$|\Phi\rangle$ in the distinctive CCM exponentiated forms,
\begin{equation}
|\Psi\rangle=e^{S}|\Phi\rangle\,; \quad \langle\tilde{\Psi}|=\langle\Phi|\tilde{S}e^{-S}\,.  \label{exp_para}
\end{equation}
In principle the correlation operator $\tilde{S}$ may be expressed in
terms of its counterpart $S$ as
\begin{equation}
\langle\Phi|\tilde{S} = \frac{\langle\Phi|e^{S^{\dagger}}e^{S}}{\langle\Phi|e^{S^{\dagger}}e^{S}|\Phi\rangle}\,,  \label{correlation-opererators-relationship}
\end{equation}
using Hermiticity.  In practice, however, the CCM methodology chooses
not to impose this constraint.  Rather, the two correlation operators
are formally decomposed independently as
\begin{equation}
S=\sum_{I\neq 0}{\cal S}_{I}C^{+}_{I}\,; \quad \tilde{S}=1+\sum_{I\neq 0}\tilde{{\cal S}}_{I}C^{-}_{I}\,,   \label{CORRELATN_OP}
\end{equation}
where $C^{+}_{0}\equiv 1$ is defined to be the identity operator in
the respective many-body Hilbert space, and where the set index $I$
denotes a complete set of single-body configurations for all $N$
particles.  More specifically, we require that $|\Phi\rangle$ is a
fiducial (or cyclic) vector with respect to the complete set of
mutually commuting, multiconfigurational creation operators
$\{C^{+}_{I}\}$.  In other words, the set of states
$\{C^{+}_{I}|\Phi\rangle\}$ is a complete basis for the ket-state
Hilbert space.  Furthermore, $|\Phi\rangle$ is a generalized vacuum
with respect to the operators $\{C^{+}_{I}\}$, in the sense that
\begin{equation}
\langle\Phi|C^{+}_{I} = 0 = C^{-}_{I}|\Phi\rangle\,, \quad \forall I
\neq 0\,,  \label{creat-destruct-operators-relationship}
\end{equation}
where $C^{-}_{I} \equiv (C^{+}_{I})^{\dagger}$ are the corresponding
multiconfigurational destruction operators.

The rather general CCM paramerizations of Eqs.\ (\ref{exp_para}),
(\ref{CORRELATN_OP}) and (\ref{creat-destruct-operators-relationship})
lie at the heart of the CCM, and have several immediate
consequences.  A seeming drawback is that Hermiticity is not made
explicit via Eq.\ (\ref{correlation-opererators-relationship}).  Thus,
while the exact correlation operators of Eq.\ (\ref{CORRELATN_OP})
will certainly fulfill Eq.\
(\ref{correlation-opererators-relationship}), when approximations are
made (e.g., by truncating the expansions over configurations $I$ in
Eq.\ (\ref{CORRELATN_OP}), as is usually necessary in practice)
Hermiticity may be only maintained approximately.  Nevertheless, this
possible disadvantage is usually far outweighed in practice by several
advantages that flow from the CCM parametrization scheme.  Thus, for
example, it guarantees the exact preservation of the Goldstone
linked-cluster theorem, even when approximate truncations are made to
the sums in Eq.\ (\ref{CORRELATN_OP}), as we describe more fully
below.  This feature then guarantees the size-extensivity of the CCM
at any such level of approximate implementation, so that all extensive
variables, such as the GS energy, for example, scale linearly with
$N$.  For this reason, the CCM has the first important advantage that
we may work from the very outset in the thermodynamic limit ($N
\rightarrow \infty$), thereby obviating the need for any finite-size
scaling of the numerical results, as is required in many competing
methods such as the ED method.  The exponentiated CCM parametrizations
of Eq.\ (\ref{exp_para}), correspondingly lead to the second key
advantage of the method that it also exactly preserves the very
important Hellmann-Feynman theorem at any similar level of
approximation (or truncation).

In the CCM all GS information of the system is encoded in the $c$-number correlation
coefficients $\{{\cal S}_{I}, \tilde{{\cal S}}_{I}\}$.  They are themselves now found by minimization of the GS energy functional,
\begin{equation}
\bar{H}=\bar{H}[{\cal S}_{I},{\tilde{\cal S}_{I}}] \equiv
\langle\Phi|\tilde{S}e^{-S}He^{S}|\Phi\rangle\,,  \label{eq_GS_E_xpect_funct}
\end{equation}
from Eq.\ (\ref{exp_para}), with respect to each of the coefficients
$\{{\cal S}_{I},{\tilde{\cal S}}_{I}\,; \forall I \neq 0\}$
separately.  Thus, variation of $\bar{H}$ from Eq.\
(\ref{eq_GS_E_xpect_funct}) with respect to the coefficient
${\tilde{\cal S}}_{I}$, yields the condition
\begin{equation}
\langle\Phi|C^{-}_{I}e^{-S}He^{S}|\Phi\rangle = 0\,, \quad \forall I \neq 0\,,  \label{ket_eq}
\end{equation}
which is simply a coupled set of {\it nonlinear} equations for the set
of coefficients $\{{\cal S}_{I},\,\forall I \neq 0\}$, with the same
number of equations as unknown parameters.  Similarly, variation of
$\bar{H}$ from Eq.\ (\ref{eq_GS_E_xpect_funct}) with respect to the
coefficient ${\cal S}_{I}$ yields the condition
\begin{equation}
\langle\Phi|\tilde{S}e^{-S}[H,C^{+}_{I}]e^{S}|\Phi\rangle=0\,, \quad \forall I \neq 0\,,  \label{bra_eq}
\end{equation}
which is, correspondingly, a coupled set of {\it linear} equations for
the coefficients $\{{\tilde{\cal S}}_{I},\,\forall I \neq 0\}$, again
with the same number of equations as unknown parameters, once the
coefficients $\{{\cal S}_{I},\,\forall I \neq 0\}$ are used as input
after having been obtained from solving Eq.\ (\ref{ket_eq}).

The value of $\bar{H}$ from Eq.\ (\ref{eq_GS_E_xpect_funct}) at the
extremum so obtained is thus the GS energy $E$, which is hence simply
given, using Eqs.\ (\ref{CORRELATN_OP}),
(\ref{creat-destruct-operators-relationship}) and (\ref{ket_eq}), as
\begin{equation}
E=\langle\Phi|e^{-S}He^{S}|\Phi\rangle=\langle\Phi|He^{S}|\Phi\rangle\,, \label{eq_GS_E}
\end{equation}
in terms of the correlation coefficients $\{{\cal S}_{I}\}$ alone.
Clearly, the GS expectation value of any other physical operator
(e.g., the sublattice magnetization, $M$) requires a knowledge of both
sets of correlation coefficients, $\{{\cal S}_{I}\}$ and
$\{\tilde{\cal S}_{I}\}$.  We note, too, that use of Eq.\
(\ref{eq_GS_E}) in Eq.\ (\ref{bra_eq}) leads to the equivalent set of
linear equations,
\begin{equation}
\langle\Phi|\tilde{S}(e^{-S}He^{S}-E)C^{+}_{I}|\Phi\rangle=0\,, \quad \forall I \neq 0\,,  \label{bra_eq_alternative}
\end{equation}
for the coefficients $\{\tilde{\cal S}_{I},\,\forall I \neq 0\}$.
Equation (\ref{bra_eq_alternative}) is just a set of generalized
linear eigenvalue equations for these coefficients.

So far no approximations have yet been made in the CCM procedure and
implementation.  It is clear, however, that Eq.\ (\ref{ket_eq}),
which determines the set of creation coefficients $\{{\cal
  S}_{I},\,\forall I \neq 0\}$, is intrinsically highly nonlinear in
view of the exponential terms, $e^{\pm S}$, and one may wonder if
approximations are needed in practice to truncate their
infinite-series expansions.  We note, however, that the (exponentiated
forms of the) operator $S$ only ever enter the equations to be solved
[i.e., Eqs.\ (\ref{ket_eq}) and (\ref{bra_eq_alternative})] in the
combination $e^{-S}He^{S}$ of a similarity transformation of the
Hamiltonian.  This may be expanded in terms of the well-known nested
commutator series,
\begin{equation}
e^{-S}He^{S} = \sum^{\infty}_{n=0}\frac{1}{n!}[H,S]_{n}\,,  \label{eq_expon_nested_commutator}
\end{equation}
where $[H,S]_{n}$ is an $n$-fold nested commutator, defined
iteratively as
\begin{equation}
[H,S]_{n}=[[H,S]_{n-1},S]\,; \quad [H,S]_{0}=H\,.
\end{equation}
A further key feature of the CCM is that this otherwise infinite sum
in Eq.\ (\ref{eq_expon_nested_commutator}) now (usually, as here)
terminates exactly at some low, finite order, when used in the
equations to be solved.  The reasons are that all of the terms
in the expansion of Eq.\ (\ref{CORRELATN_OP}) for $S$ commute with
one another, and also that $H$ itself (usually, as here) is of finite order
in the relevant single-particle operators.

Thus, for example, if $H$ contains up to $m$-body interactions, in its
second-quantized form it will comprise sums of terms involving
products of up to $2m$ one-body destruction and creation operators.
In this case the sum in Eq.\ (\ref{eq_expon_nested_commutator})
terminates exactly at the term with $n=2m$.  In our present case,
where the Hamiltonian of Eq.\ (\ref{eq_H}) is bilinear in the SU(2)
spin operators, the sum terminates at $n=2$.  We note too that the
fact that all of the operators in the set $\{C^{+}_{I}\}$ that
comprise $S$ via Eq.\ (\ref{CORRELATN_OP}) commute with each other,
automatically implies that all non-vanishing terms in the expansion in
Eq.\ (\ref{eq_expon_nested_commutator}) are linked to the Hamiltonian.
Unlinked terms simply cannot be generated, thereby guaranteeing that
the Goldstone theorem and the consequent size-extensivity are
preserved, even when truncations are made for the correlation
operators $S$ and $\tilde{S}$.

Hence, the {\it only} approximation that is ever made in practice
to implement the CCM is to restrict the set of multiconfigurational
indices $\{I\}$ that we retain in the expansions of $S$ and
$\tilde{S}$ in Eq.\ (\ref{CORRELATN_OP}) to some suitable (finite
or infinite) subset.  The choice of both model state $|\Phi\rangle$
and the indices $\{I\}$ retained must clearly be made on physical
grounds.  Hence, we now turn to how such choices are made for
spin-lattice models in general, and for the specific system under
present study in particular.

For a general quantum spin-lattice problem, the simplest choice of
model state $|\Phi\rangle$ is an independent-spin product state in
which the spin projection of the spin on each lattice site, along some
specified quantization axis, is chosen independently.  Clearly, the
two-sublattice, collinear N\'{e}el AFM state shown in Fig.\
\ref{model_pattern}(a) is precisely of this form, as are other similar
(quasi-)classical states with perfect magnetic LRO.  In order to treat
all such states in a universal fashion, it is highly convenient to
make a passive rotation of each spin independently (i.e., by choosing
local spin quantization axes on each site independently), so that every
spin on every site points downwards, say, in the negative $z_{s}$
direction, as shown in Fig.\ \ref{model_pattern}.  Such rotations
are unitary transformations that preserve the basic SU(2) commutation
relations of Eq.\ (\ref{SU2_comm_relatn}).  Hence every lattice site
$k$ is completely equivalent to all others, whatever the choice of
such an independent-spin product, quasiclassical model spin state
$|\Phi\rangle$, all of which now take the universal form
$|\Phi\rangle=|$$\downarrow\downarrow\downarrow\cdots\downarrow\rangle$
in their own choices of local spin-coordinate frames for each site $k$
separately.

It is clear that $|\Phi\rangle$ so defined can now be taken as a
fiducial vector with respect to a set of mutually commuting creation
operators $\{C^{+}_{I}\}$, which are hence now chosen as a product of
single-spin raising operators, $s^{+}_{k} \equiv
s^{x}_{k}+is^{y}_{k}$.  Thus, $C^{+}_{I} \rightarrow
s^{+}_{k_{1}}s^{+}_{k_{2}}\cdots s^{+}_{k_{n}};\; n=1,2,\cdots , 2sN$,
and the set index $I$ thus becomes a set of lattice-site indices, $I
\rightarrow \{k_{1},k_{2},\cdots , k_{n};\; n=1,2,\cdots , 2sN\}$, in
which each site index may appear up to $2s$ times at most.  Once
the local spin coordinates have been chosen for the given model state
$|\Phi\rangle$, as specified above, one needs simply to re-express the
Hamiltonian $H$ in terms of them.

Our approximations now clearly involve simply a choice of which
configurations $\{I\}$ to retain in the decompositions of Eq.\
(\ref{CORRELATN_OP}) for the CCM correlation operators
$(S,\tilde{S})$, in terms of which all GS quantities may be expressed.
A rather general such approximation scheme is the so-called
SUB$n$--$m$ scheme, which has proven to be extremely powerful in
practice for a wide variety of applications to spin-lattice problems
ranging from unfrustrated to highly frustrated models.  It retains,
for given values of the two truncation indices $n$ and $m$, all
multi-spin configurations involving a maximum of $n$ spin-flips (where
each spin flip requires the action of a spin-raising operator
$s^{+}_{k}$ acting once) that span a range of up to $m$ contiguous
sites at most.  A set of lattice sites is defined to be contiguous for
these purposes if every site in the set is the NN of at least one
other in the set (in a specified geometry).  Evidently, as both
truncation indices $n$ and $m$ become indefinitely large, the
approximation becomes exact, and different sub-schemes can be specified
according to how each index approaches the exact infinite limit.

If we first let $m \rightarrow \infty$, for example, we have the
so-called SUB$n$ $\equiv$ SUB$n$--$\infty$ scheme, which is just the
CCM truncation scheme employed rather generically for systems defined
in a spatial continuum (rather than on a discrete lattice, as here).
Examples to which the SUB$n$ scheme have been extensively applied,
include atoms and molecules in quantum chemistry
\cite{Bartlett:1989_ccm}, finite atomic nuclei or infinite nuclear
matter in nuclear physics \cite{Kummel:1978_ccm} (and see, e.g.,
Refs.\
\cite{Bishop:1991_TheorChimActa_QMBT,Bishop:1978_ccm,Bishop:1982_ccm}
for further details).  By contrast to continuum theories, for which
the notion of contiguity is not readily applicable, in lattice
theories both indices $n$ and $m$ may be kept finite.  In this case a
very widely used scheme is the so-called LSUB$m$ scheme
\cite{Fa:2004_QM-coll,Zeng:1998_SqLatt_TrianLatt}, which is defined to
retain, at the $m$th level of approximation, all spin clusters
described by multi-spin configurations in the index set $\{I\}$ that
are defined over any possible lattice animal (or equivalently,
polyomino) of maximal size $m$ on the lattice.  A lattice animal is
defined here, in the usual graph-theoretic sense, to be a configured
set of contiguous (in the above sense) lattice sites.  Clearly, the
LSUB$m$ scheme is equivalent to the previous SUB$n$--$m$ scheme when
$n=2sm$ for particles of spin quantum number $s$, i.e., LSUB$m \equiv$
SUB$2sm$--$m$.  Just this LSUB$m$ scheme was what was employed in our
earlier studies of spin-$\frac{1}{2}$ honeycomb lattice models
\cite{Bishop:2015_honey_low-E-param,DJJF:2011_honeycomb}, for example.

However, the number $N_{f}=N_{f}(m)$ of fundamental spin configurations that
are distinct under the symmetries of the lattice and the specified
model state $|\Phi\rangle$ (i.e., the effective size of the index set
$\{I\}$), and which are retained at a given $m$th level of LSUB$m$
approximation, is lowest for $s=\frac{1}{2}$ and rises sharply as $s$
is increased.  Since $N_{f}(m)$ also typically rises
super-exponentially with the truncation index $m$, an alternative
scheme for models with $s > \frac{1}{2}$ is often preferable.  One such
alternative is to set $m=n$ and employ the ensuing SUB$n$--$n$ scheme,
as we shall do here.  Clearly LSUB$m$ $\equiv$ SUB$m$--$m$ only in the
special case $s=\frac{1}{2}$.  For $s>\frac{1}{2}$ we have SUB$n$--$n$
$\subset$ LSUB$n$.  Just as for the LSUB$n$ scheme, however, the
number $N_{f}$ of fundamental configurations retained at a given $n$th
level of approximation, also rises as the spin quantum number $s$
is increased.  Thus, for example, whereas for the $s=\frac{1}{2}$
honeycomb-lattice HAF the highest LSUB$m$ approximation attainable
with available supercomputing power using the N\'{e}el state as CCM
model state \cite{Bishop:2015_honey_low-E-param,DJJF:2011_honeycomb}
was $m=12$ (for which $N_{f}=103,097$), we are now constrained to
SUB$n$--$n$ approximations with $n \leq 10$ for the cases $1 \leq s
\leq \frac{9}{2}$ considered here.  Thus, for example, at the
SUB10--10 level of approximation with the N\'{e}el model state, we
have $N_{f}=219,521$ for the case $s=1$, and $N_{f}=538,570$ for the
case $s=\frac{9}{2}$.

In order to derive and then solve \cite{Zeng:1998_SqLatt_TrianLatt}
the corresponding sets of CCM equations for the correlation
coefficients $\{{\cal S}_{I}, \tilde{{\cal S}}_{I}\}$ we employ
massively parallel computing \cite{ccm_code}.  Once these coefficients
have been obtained at a given SUB$n$--$n$ level of truncation we may
calculate any GS property of the system at the same level of
approximation.  For example, we may calculate the order parameter, as
defined in Eq.\ (\ref{stag_M}).  In terms of the local rotated
spin-coordinate frames that we have described above, it takes the simple
form,
\begin{equation}
M = -\frac{1}{N}\sum^{N}_{k=1}\langle\Phi|\tilde{S}
  e^{-S}s^{z}_{k}e^{S}|\Phi\rangle\,.   \label{M_eq}
\end{equation}

The final step now involves the sole approximation made in our
entire CCM procedure, viz., the extrapolation of the ``raw'' SUB$n$--$n$
data points for our calculated low-energy parameters to the exact $n \rightarrow \infty$ limit.
While no exact extrapolation rules are known, a large body of experience
has by now been accumulated from many applications that have been made of the method to a large variety of spin-lattice models.  For example, a very well tested and highly accurate extrapolation ansatz for the GS energy per spin has been shown to be 
(and see, e.g., Refs.\
\cite{Fa:2004_QM-coll,DJJF:2011_honeycomb,Bishop:2000_XXZ,Kruger:2000_JJprime,Fa:2001_SqLatt_s1,Darradi:2005_Shastry-Sutherland,Darradi:2008_J1J2mod,Bi:2008_EPL_J1J1primeJ2_s1,Bi:2008_JPCM_J1xxzJ2xxz_s1,Bi:2009_SqTriangle,Bishop:2010_UJack,Bishop:2010_KagomeSq,Bishop:2011_UJack_GrtSpins,PHYLi:2012_SqTriangle_grtSpins,PHYLi:2012_honeycomb_J1neg,Bishop:2012_honeyJ1-J2,Bishop:2012_honey_circle-phase,Li:2012_honey_full,Li:2012_anisotropic_kagomeSq,RFB:2013_hcomb_SDVBC})
\begin{equation}
\frac{E(n)}{N} = e_{0}+e_{1}n^{-2}+e_{2}n^{-4}\,.     \label{extrapo_E}
\end{equation}

Unsurprisingly, all other GS quantities are found to converge less rapidly than the GS energy, as the truncation index $n$ is increased (i.e., with leading exponents less than two).  Thus, for example, for unfrustrated models as considered here, a scaling ansatz for the magnetic order parameter, $M(n)$, with leading power $1/n$ (rather than $1/n^{2}$ as for the GS energy),
\begin{equation}
M(n) = m_{0}+m_{1}n^{-1}+m_{2}n^{-2}\,,   \label{M_extrapo_standard}
\end{equation}
has been found to fit the CCM data points extremely well (and see, e.g., Refs.\
\cite{Bishop:2000_XXZ,Kruger:2000_JJprime,Fa:2001_SqLatt_s1,Darradi:2005_Shastry-Sutherland,Bi:2009_SqTriangle,Bishop:2010_UJack,Bishop:2010_KagomeSq,Bishop:2011_UJack_GrtSpins,PHYLi:2012_SqTriangle_grtSpins,PHYLi:2012_honeycomb_J1neg,Bishop:2012_honeyJ1-J2,Bishop:2012_honey_circle-phase,RFB:2013_hcomb_SDVBC}).
Similar schemes, with the same leading exponent as for $M$, have
also been successfully used previously for both the spin stiffness
$\rho_{s}$ \cite{Bishop:2015_honey_low-E-param,Darradi:2008_J1J2mod,SEKruger:2006_spinStiff,Gotze:2016_triang},
\begin{equation}
\rho_{s}(n) = s_{0}+s_{1}n^{-1}+s_{2}n^{-2}\,,   \label{Eq_sstiff}
\end{equation}
and the zero-field magnetic susceptibility, $\chi$
\cite{Bishop:2015_honey_low-E-param,Darradi:2008_J1J2mod,Farnell:2009_Xcpty_ExtMagField,Gotze:2016_triang},
\begin{equation}
\chi(n) = x_{0}+x_{1}n^{-1}+x_{2}n^{-2}\,.   \label{Eq_X}
\end{equation}

Since each of the extrapolation schemes of Eqs.\
(\ref{extrapo_E})--(\ref{Eq_X}) contains three fitting parameters, it
is obviously preferable to use at least four SUB$n$--$n$ data points
in order to obtain stable and robust fits.  Furthermore since the
lowest-order SUB2--2 approximants are less likely to conform well to
the large-$n$ limiting forms, all of our fits for $E/N$ and $M$ to
Eqs.\ (\ref{extrapo_E}) and (\ref{M_extrapo_standard}) are performed
using SUB$n$--$n$ data points with $n=\{4,6,8,10\}$.  Whereas we are
able to perform SUB$n$--$n$ calculations for the cases $s \leq
\frac{9}{2}$ for the honeycomb-lattice HAF with $n \leq 10$ based on
the N\'{e}el state of Fig.\ \ref{model_pattern}(a) as CCM model state,
the reduced symmetry of both the twisted N\'{e}el state of Fig.\
\ref{model_pattern}(b) and the canted N\'{e}el state of Fig.\
\ref{model_pattern}(c) restricts our corresponding SUB$n$--$n$
calculations of both $\rho_{s}$ and $\chi$ to $n \leq 8$.  Thus, for
example, at the SUB8--8 level of approximation with the twisted
N\'{e}el state of Fig.\ \ref{model_pattern}(b) as model state, we have
a number of fundamental configurations $N_{f}=352,515$ for the case
$s=1$, and $N_{f}=753,729$ for the case $s=\frac{9}{2}$.  The
corresponding SUB8--8 numbers with the canted N\'{e}el state of Fig.\
\ref{model_pattern}(c) as model state are $N_{f}=59,517$ for the case
$s=1$, and $N_{f}=127,239$ for the case $s=\frac{9}{2}$.  For the
extrapolations for $\rho_{s}$ and $\chi$ from Eqs.\ (\ref{Eq_sstiff})
and (\ref{Eq_X}) our results shown below are from fits using
SUB$n$--$n$ data points with $n=\{4,6,8\}$.  Their stability and
robustness has been demonstrated, however, by comparison in each case
with corresponding fits using data sets with $n=\{2,4,6,8\}$.

\section{RESULTS}
\label{results_sec}
We show in Table \ref{tbl_low-E-param} our extrapolated set of
low-energy parameters for the HAF on the honeycomb lattice for all
values of the spin quantum number in the range $\frac{1}{2} \leq s
\leq \frac{9}{2}$.
\begin{table}[t]   
  \caption{GS parameters of the HAF on the honeycomb lattice, with NN interactions only (of strength $J_{1}>0$), for various values of the spin quantum number $s$.  Results for the GS energy per spin $E/N$ and magnetic order parameter $M$ are extrapolations using CCM SUB$n$--$n$ results with $n=\{4,6,8,10\}$ fitted to Eqs.\ (\ref{extrapo_E}) and (\ref{M_extrapo_standard}), respectively, while those for the spin stiffness $\rho_{s}$ and the zero-field transverse magnetic susceptibility $\chi$ are corresponding extrapolations with $n=\{4,6,8\}$ fitted to Eqs.\ (\ref{Eq_sstiff}) and (\ref{Eq_X}), respectively.}  
\vspace{0.2cm}
\begin{center}    
\begin{tabular}{ccccc}   \hline\hline \\ [-1.5ex]   
$s$ & $E/(NJ_{1}s^{2})$ & $M/s$ & $\rho_{s}/(J_{1}d^{2}s^{2})$ & $J_{1}\chi$ \\ [0.5ex] \hline \\ [-1.7ex]
$\frac{1}{2}$ & -2.17866 & 0.5459 & 0.5293 & 0.0852 \\ [0.5ex]
1 & -1.83061 & 0.7412 & 0.6208 & 0.1165 \\  [0.5ex]
$\frac{3}{2}$ & -1.71721 & 0.8249 & 0.6647 & 0.1287 \\ [0.5ex]
2 & -1.66159 & 0.8689 &  0.6874 & 0.1376 \\ [0.5ex]
$\frac{5}{2}$ & -1.62862 & 0.8955 & 0.7008 & 0.1433 \\ [0.5ex] 
3 & -1.60681 & 0.9132  &  0.7095 & 0.1471 \\ [0.5ex]
$\frac{7}{2}$ & -1.59133 & 0.9258  & 0.7156 &  0.1499 \\ [0.5ex]
4 & -1.57976 & 0.9351 & 0.7201 & 0.1522 \\ [0.5ex]
$\frac{9}{2}$ & -1.57080  & 0.9424  & 0.7236 & 0.1538 \\ [0.5ex] \hline \\ [-1.7ex]
$\infty$ & -1.5 & 1 & 0.75 & 0.1667 \\ [0.5ex] \hline\hline
\end{tabular}
\end{center}
\label{tbl_low-E-param}
\end{table}
The extrapolated ($n \rightarrow \infty$) values $e_{0}$ and $m_{0}$
for the GS energy per spin $E/N$ and magnetic order parameter $M$ from
Eqs.\ (\ref{extrapo_E}) and (\ref{M_extrapo_standard}), respectively,
are obtained using fits to our calculated CCM SUB$n$--$n$
approximants with $n=\{4,6,8,10\}$.  The corresponding extrapolated ($n
\rightarrow \infty$) values $s_{0}$ and $x_{0}$ for the spin stiffness
$\rho_{s}$ and the zero-field transverse magnetic susceptibility
$\chi$ from Eqs.\ (\ref{Eq_sstiff}) and (\ref{Eq_X}), respectively,
are obtained using fits to our calculated SUB$n$--$n$ approximants with
$n=\{4,6,8\}$.  An indication of the errors inherent in the fits can
be obtained for the particular case $s=\frac{1}{2}$, for which it is
possible to perform SUB$n$--$n$ approximations with higher values of
$n$ than for the cases $s>\frac{1}{2}$, due to the significantly
reduced number of fundamental configurations $N_{f}$ for each quantity
in this specific case.  Thus, for example, the results
\cite{Bishop:2015_honey_low-E-param} for the case $s=\frac{1}{2}$
using SUB$n$--$n$ approximants with $n=\{6,8,10,12\}$ are
$E/(NJ_{1}s^{2})=-2.17864$, $M/s=0.5428$, and $J_{1}\chi=0.0847$,
while the corresponding result for $\rho_{s}$ using SUB$n$--$n$
approximants with $n=\{6,8,10\}$ is
$\rho_{s}/(J_{1}d^{2}s^{2})=0.5296$.  All of these are in remarkably
close agreement with those shown in Table \ref{tbl_low-E-param}, where
the fits have been made using SUB$n$--$n$ approximants of lower orders
in each case.

We note that as an indicator of the accuracy of our CCM results, we
have already made a detailed comparison in an earlier paper
\cite{Bishop:2015_honey_low-E-param} of our results for the low-energy
parameters with the corresponding largest-scale and numerically most
accurate QMC results available for the isotropic, honeycomb-lattice
HAF, namely for the spin-$\frac{1}{2}$ case
\cite{Low:2009_honey,Jiang:2012_honey}.  This is the extreme quantum
limit, where one expects the effects of quantum correlations to be
greatest.  For example, our CCM result
\cite{Bishop:2015_honey_low-E-param} for the GS energy of the
spin-$\frac{1}{2}$ model is $E/(NJ_{1})=-0.54466(2)$, while the
corresponding best available QMC result \cite{Low:2009_honey} is
$E/(NJ_{1})=-0.54455(20)$.  The interested reader is referred
specifically to Table I of Ref.\ \cite{Bishop:2015_honey_low-E-param}, and the discussion surrounding it, for further details of
the corresponding agreement for other low-energy parameters of the
spin-$\frac{1}{2}$ model.  There is no reason at all why the CCM
results for the models with $s > \frac{1}{2}$ should not be at least
as accurate as those for the spin-$\frac{1}{2}$ model.

\begin{figure}[t]
\begin{center}
\mbox{
\hspace{-1cm}\subfigure[]{\includegraphics[width=6cm,angle=270]{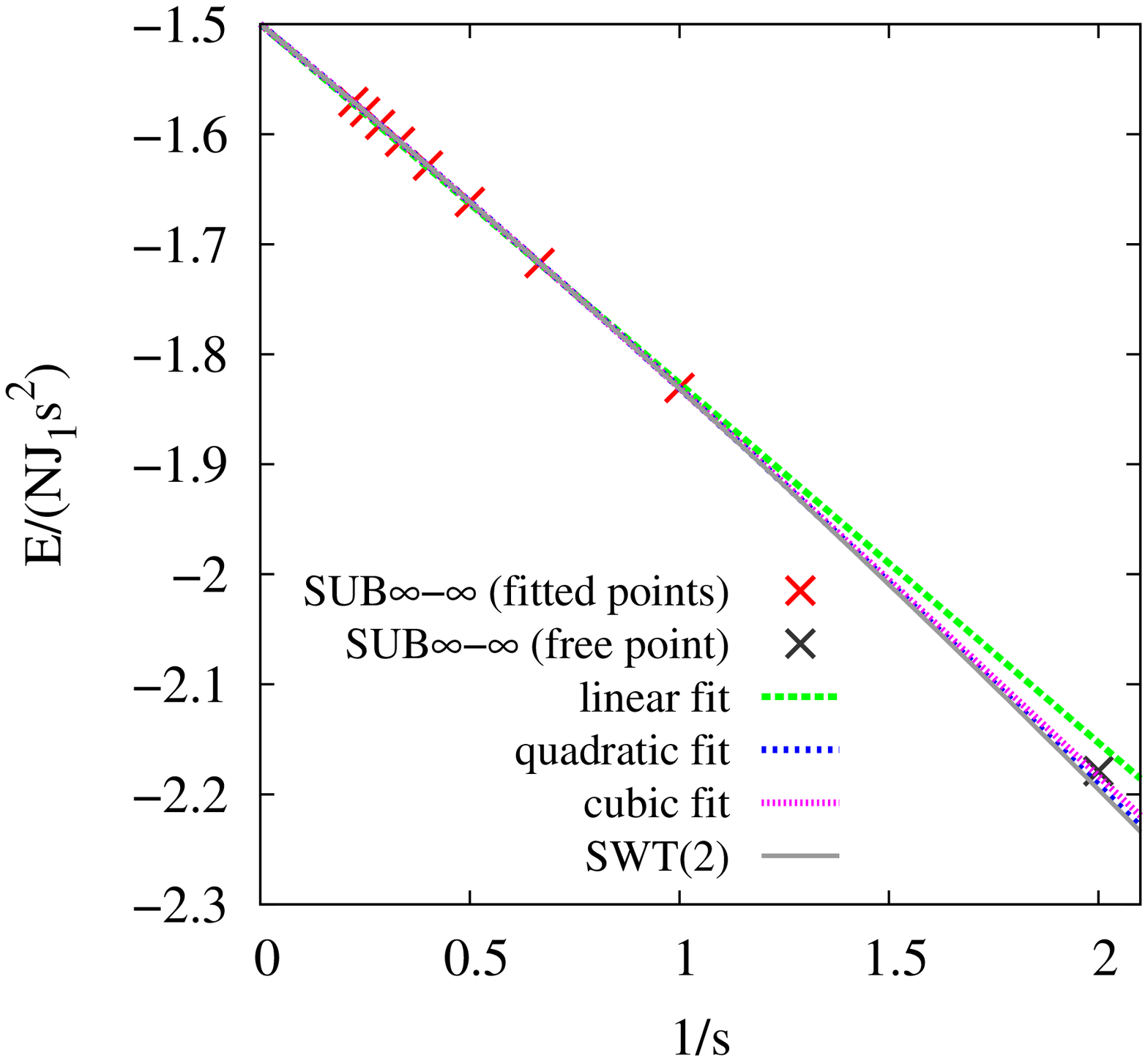}}
\hspace{-1.5cm}\subfigure[]{\includegraphics[width=6cm,angle=270]{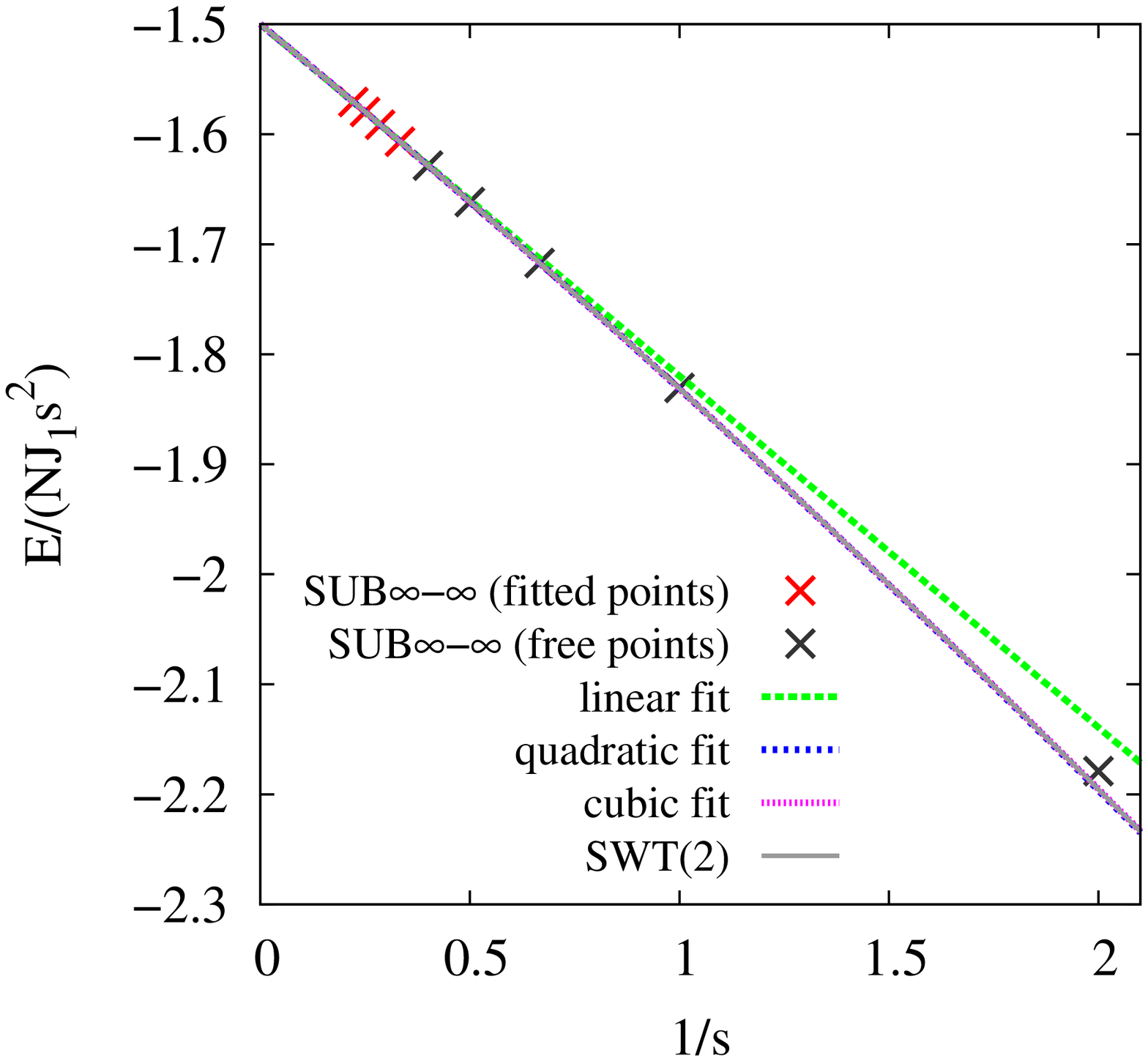}}
  }
  \caption{Extrapolated CCM results for the scaled GS energy per spin,
    $E/(NJ_{1}s^{2})$, for the honeycomb-lattice HAF with NN
    interactions only as a function of $1/s$, compared with those of
    SWT(2) \cite{Zheng:1991_honey}.  The cross ($\times$) symbols show
    the CCM data points, while the linear, quadratic and cubic fits
    are based on least-squares fits to the first two, three or four
    terms only of Eq.\ (\ref{honey-pure_E-fit_inversePower}), using
    the data points with (a)
    $s=\{1,\,\frac{3}{2},\,2,\,\frac{5}{2},\,3,\,\frac{7}{2},\,4,\,\frac{9}{2}\}$,
    and (b) $s=\{3,\,\frac{7}{2},\,4,\,\frac{9}{2}\}$.}
\label{E_multiSpins}
\end{center}
\end{figure}

\begin{figure}[t]
\begin{center}
\mbox{
\hspace{-1cm}\subfigure[]{\includegraphics[width=6cm,angle=270]{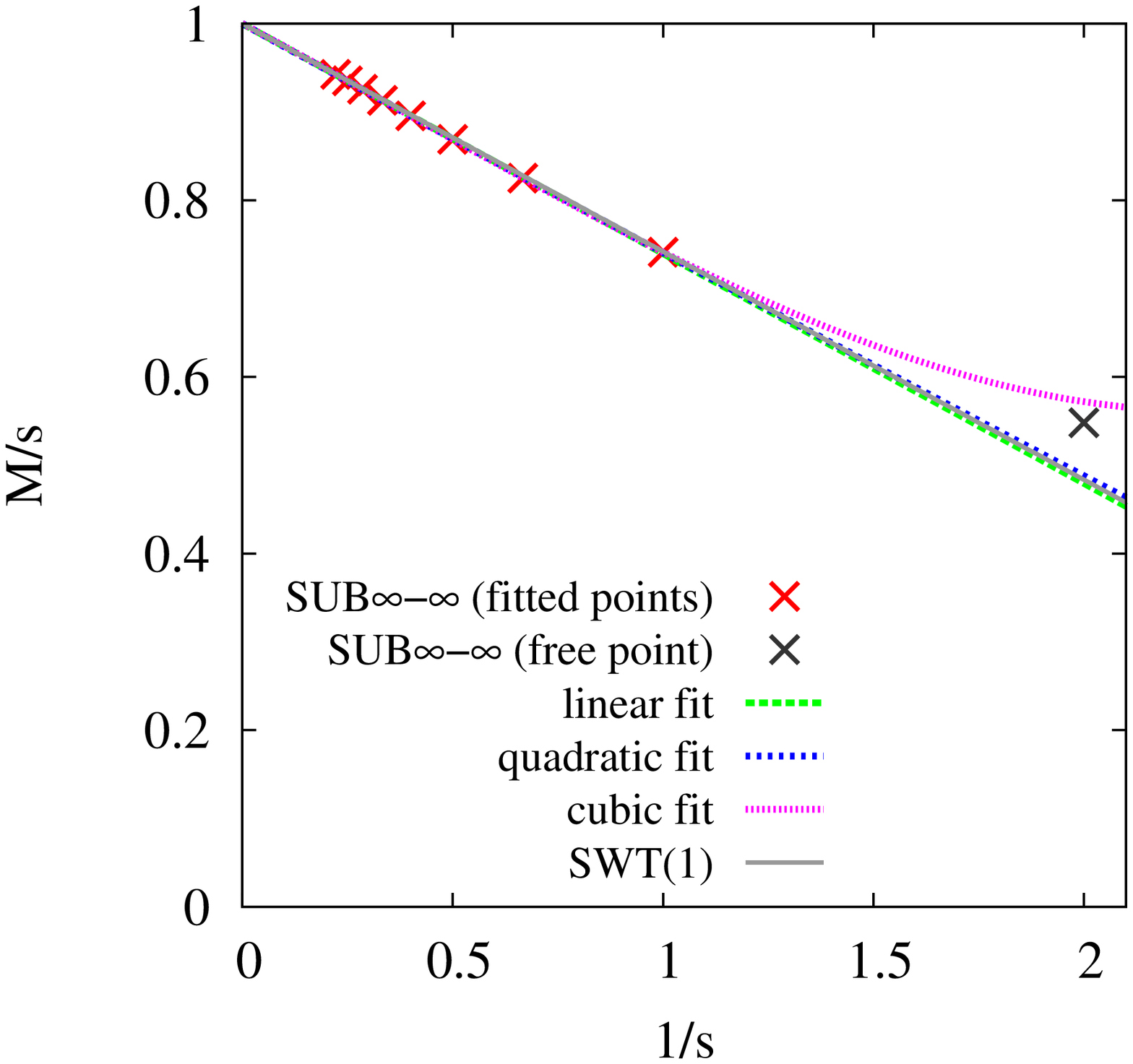}}
\hspace{-1.5cm}\subfigure[]{\includegraphics[width=6cm,angle=270]{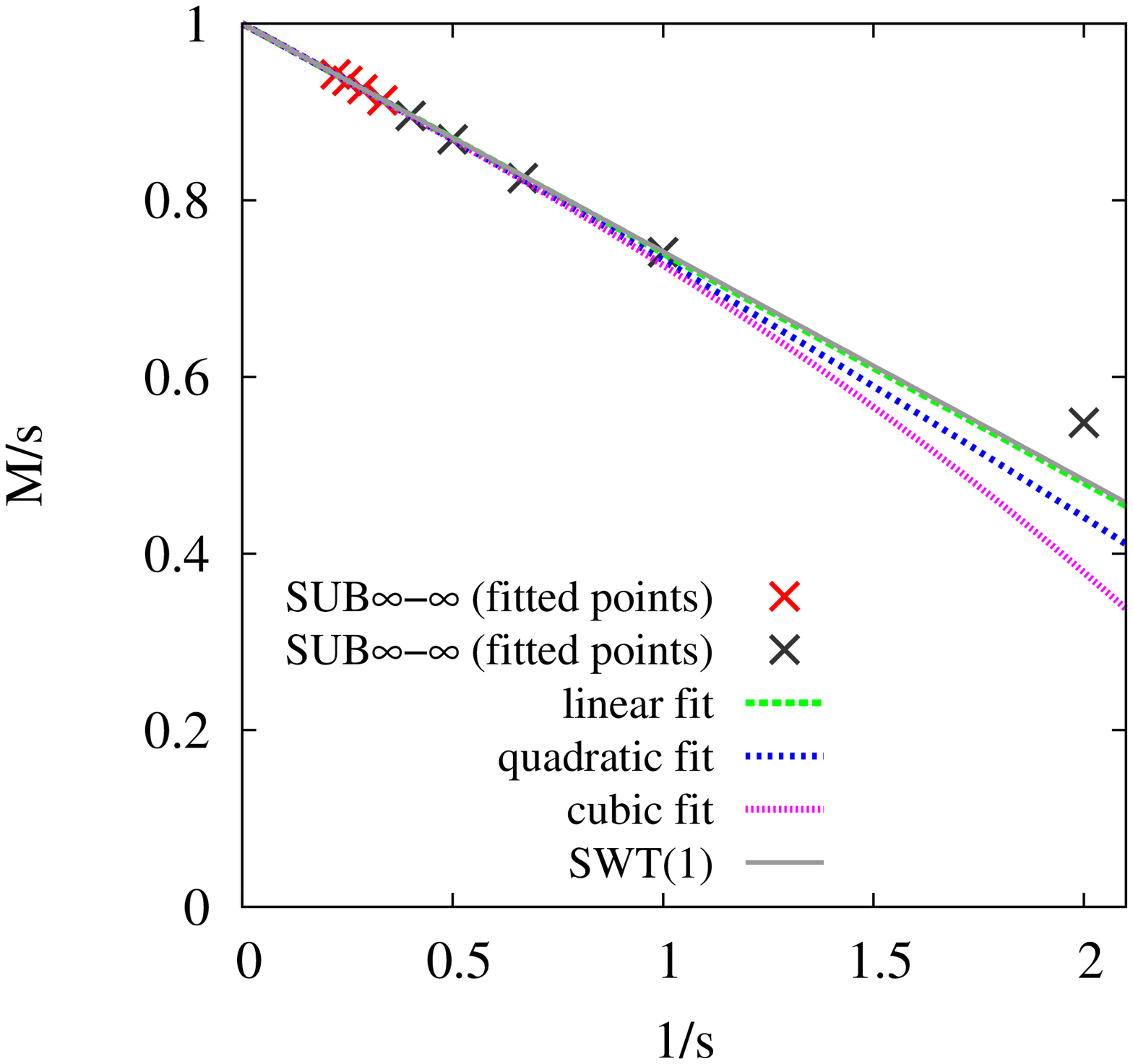}}
  }
\caption{Extrapolated CCM results for the scaled GS magnetic order parameter,
    $M/s$, for the honeycomb-lattice HAF with NN
    interactions only as a function of $1/s$, compared with those of
    SWT(1) \cite{Zheng:1991_honey}.  The cross ($\times$) symbols show
    the CCM data points, while the linear, quadratic and cubic fits
    are based on least-squares fits to the first two, three or four
    terms only of Eq.\ (\ref{honey-pure_M-fit_inversePower}), using
    the data points with (a)
    $s=\{1,\,\frac{3}{2},\,2,\,\frac{5}{2},\,3,\,\frac{7}{2},\,4,\,\frac{9}{2}\}$,
    and (b) $s=\{3,\,\frac{7}{2},\,4,\,\frac{9}{2}\}$.}
\label{M_multiSpins}
\end{center}
\end{figure}

\begin{figure}[t]
\begin{center}
\mbox{
\hspace{-1cm}\subfigure[]{\includegraphics[width=6cm,angle=270]{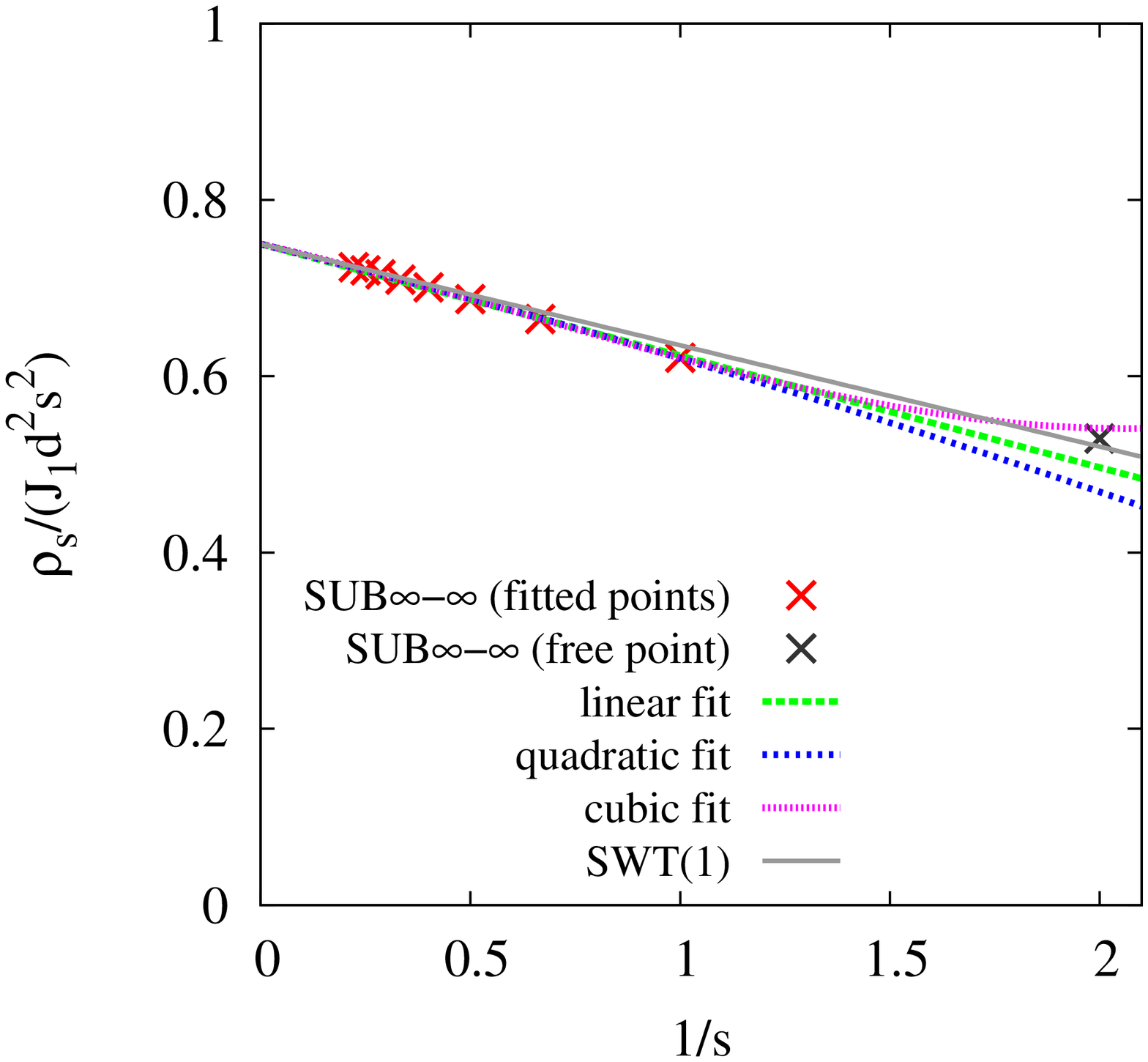}}
\hspace{-1.5cm}\subfigure[]{\includegraphics[width=6cm,angle=270]{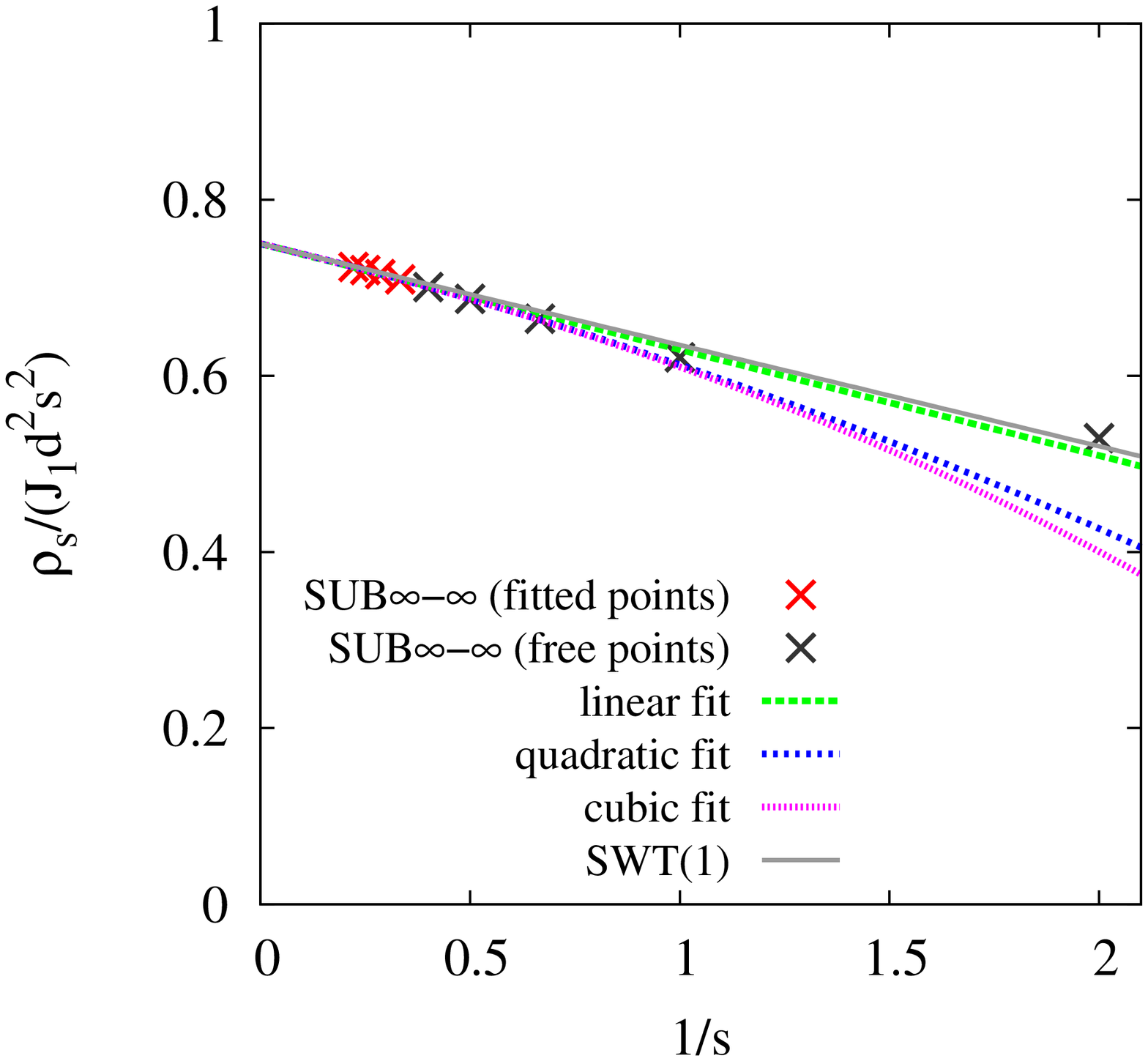}}
  }
\caption{Extrapolated CCM results for the scaled spin stiffness, $\rho_{s}/(J_{1}d^{2}s^{2})$, for the honeycomb-lattice HAF with NN
    interactions only as a function of $1/s$, compared with those of
    SWT(1) \cite{Zheng:1991_honey,Mattsson:1994_honey}.  The cross ($\times$) symbols show
    the CCM data points, while the linear, quadratic and cubic fits
    are based on least-squares fits to the first two, three or four
    terms only of Eq.\ (\ref{honey-pure_sStiff-fit_inversePower}), using
    the data points with (a)
    $s=\{1,\,\frac{3}{2},\,2,\,\frac{5}{2},\,3,\,\frac{7}{2},\,4,\,\frac{9}{2}\}$,
    and (b) $s=\{3,\,\frac{7}{2},\,4,\,\frac{9}{2}\}$.}
\label{sStiff_multiSpins}
\end{center}
\end{figure}

\begin{figure}[t]
\begin{center}
\mbox{
\hspace{-1cm}\subfigure[]{\includegraphics[width=6cm,angle=270]{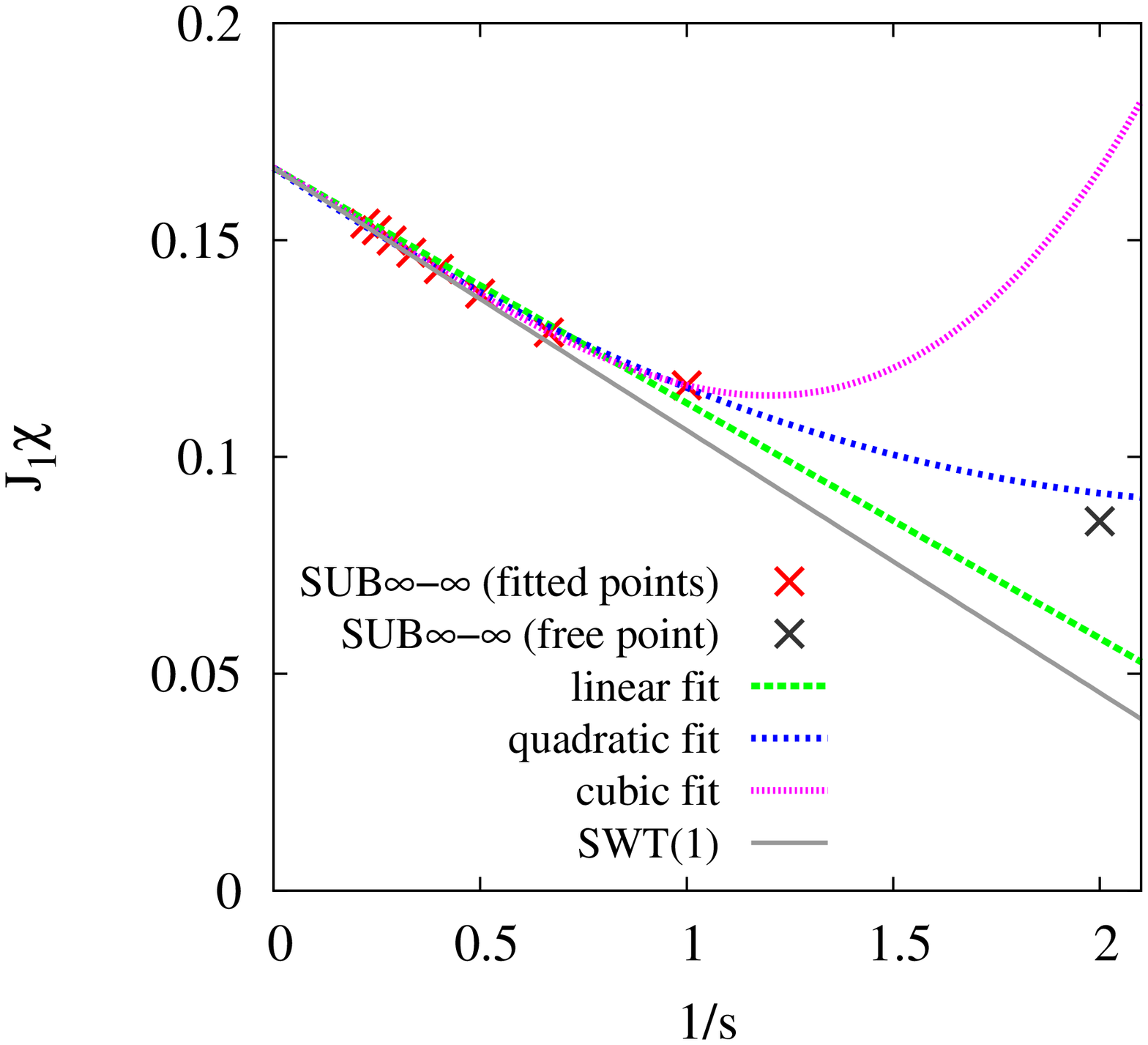}}
\hspace{-1.5cm}\subfigure[]{\includegraphics[width=6cm,angle=270]{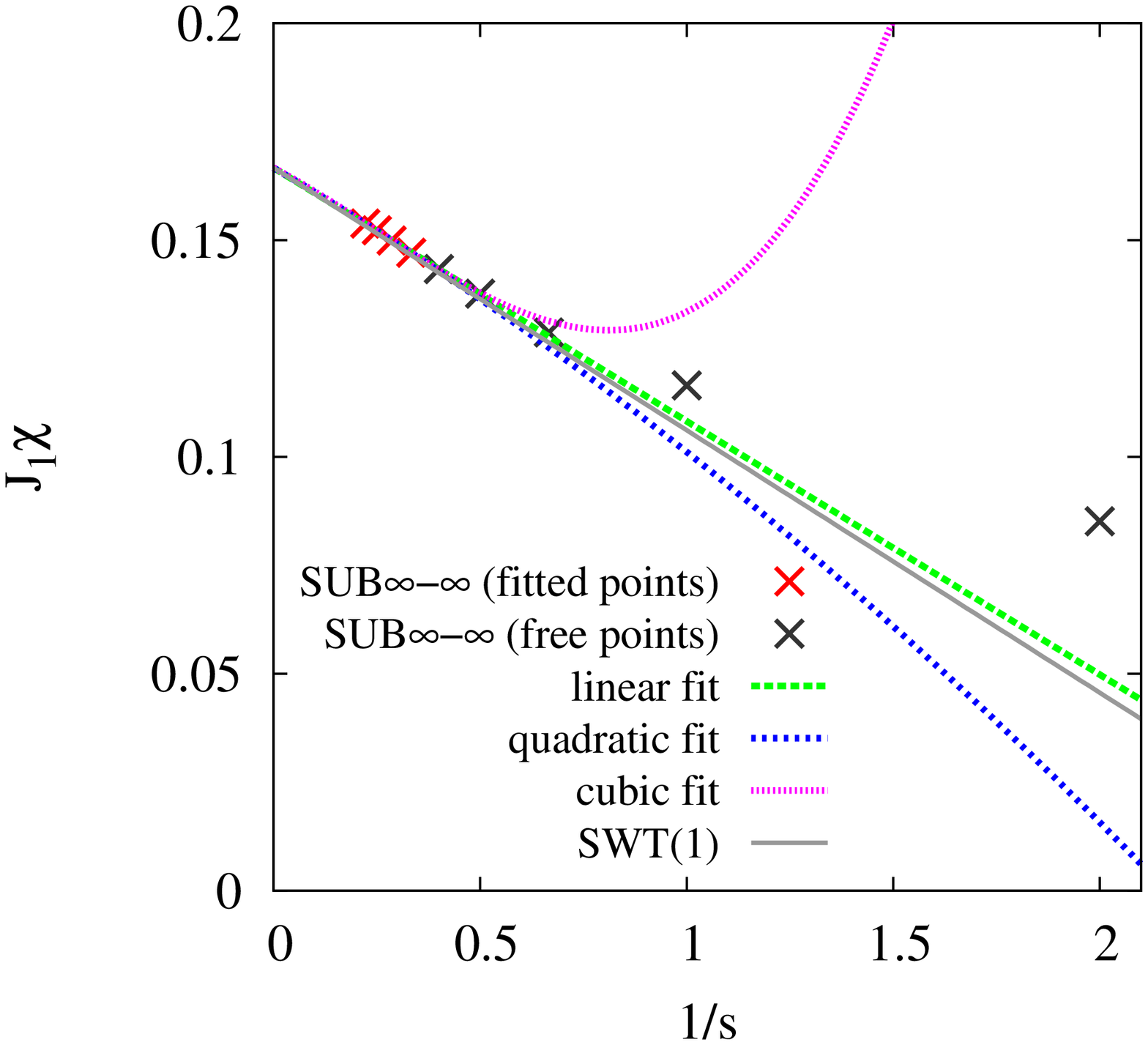}}
  }
\caption{Extrapolated CCM results for the scaled zero-field transverse magnetic susceptibility, $J_{1}\chi$, for the honeycomb-lattice HAF with NN
    interactions only as a function of $1/s$, compared with those of
    SWT(1) \cite{Zheng:1991_honey}.  The cross ($\times$) symbols show
    the CCM data points, while the linear, quadratic and cubic fits
    are based on least-squares fits to the first two, three or four
    terms only of Eq.\ (\ref{honey-pure_X-fit_inversePower}), using
    the data points with (a)
    $s=\{1,\,\frac{3}{2},\,2,\,\frac{5}{2},\,3,\,\frac{7}{2},\,4,\,\frac{9}{2}\}$,
    and (b) $s=\{3,\,\frac{7}{2},\,4,\,\frac{9}{2}\}$.}
\label{X_multiSpins}
\end{center}
\end{figure}

We can now also use our results to estimate the leading quantum corrections to the classical ($s \rightarrow \infty$) values, which we also show in the last line of Table \ref{tbl_low-E-param} [(and see Eqs.\ (\ref{sStiff_neel_classical}) and (\ref{chi_neel_classical})].  We thus develop each of the low-energy parameters as a simple power-series in $1/s$,
\begin{equation}
\frac{E(s)}{N} = J_{1}s^{2}\left(-\frac{3}{2} + \frac{\epsilon_{1}}{s} + \frac{\epsilon_{2}}{s^{2}} + \frac{\epsilon_{3}}{s^{3}} + \cdots \right) \,,  \label{honey-pure_E-fit_inversePower}
\end{equation} 
\begin{equation}
M(s) = s\left(1 + \frac{\mu_{1}}{s} + \frac{\mu_{2}}{s^{2}} + \frac{\mu_{3}}{s^{3}} + \cdots \right)\,,   \label{honey-pure_M-fit_inversePower}
\end{equation}
\begin{equation}
\rho_{s}(s) = J_{1}d^{2}s^{2}\left(\frac{3}{4} + \frac{\rho_{1}}{s} + \frac{\rho_{2}}{s^{2}} + \frac{\rho_{3}}{s^{3}} + \cdots \right)\,,   \label{honey-pure_sStiff-fit_inversePower}
\end{equation} 
\begin{equation}
\chi(s) = \frac{1}{J_{1}}\left(\frac{1}{6} + \frac{\chi_{1}}{s} + \frac{\chi_{2}}{s^{2}} + \frac{\chi_{3}}{s^{3}}+ \cdots \right)\,,   \label{honey-pure_X-fit_inversePower}
\end{equation} 
exactly as also emerges in an SWT analysis.  We show in Table \ref{tbl_low-E-param-ExtrapoFits} the values of the two leading quantum coefficients in each of these expansions, obtained in six separate least-squares fits.  
\begin{table}   
  \caption{Lowest-order coefficients in the $1/s$ expansions of the low-energy parameters for the HAF on the honeycomb lattice, as defined in Eqs.\ (\ref{honey-pure_E-fit_inversePower})--(\ref{honey-pure_X-fit_inversePower}).  The CCM fits are performed using the results for the spin-$s$ models indicated, and where the notation quadratic (cubic) indicates that we fit to forms that terminate after the first three (four) terms in the $1/s$ expansion.  Results from SWT \cite{Zheng:1991_honey,Mattsson:1994_honey} are shown for comparison.}  
\begin{center}    
{\tiny
\begin{tabular}{ccccccccc}   \hline\hline \\ [-0.7ex]   
Method & $\epsilon_{1}$ & $\epsilon_{2}$ & $\mu_{1}$ & $\mu_{2}$ & $\rho_{1}$ & $\rho_{2}$ & $\chi_{1}$ & $\chi_{2}$ \\ [1.7ex] \hline \\ [-0.7ex]
CCM : $1 \leq s \leq \frac{9}{2}$; quadratic & -0.3155 & -0.0152 & -0.2628 & +0.0034 & -0.1189 & -0.0109 & -0.0638 & +0.0132 \\ [1.7ex]
CCM : $1 \leq s \leq \frac{9}{2}$; cubic & -0.3145 & -0.0185 & -0.2531 & -0.0309 & -0.1105 & -0.0405 & -0.0552 & -0.0175  \\  [1.7ex]
CCM : $2 \leq s \leq \frac{9}{2}$; quadratic & -0.3149 & -0.0165 & -0.2568 & -0.0106 & -0.1136 & -0.0235 & -0.0586 &  +0.0006 \\ [1.7ex]
CCM : $2 \leq s \leq \frac{9}{2}$; cubic & -0.3148 & -0.0172 & -0.2560 &  -0.0157 & -0.1126 & -0.0291 & -0.0536 & -0.0275 \\  [1.7ex]
CCM : $3 \leq s \leq \frac{9}{2}$; quadratic & -0.3149 & -0.0167 & -0.2566 & -0.0114 & -0.1134 & -0.0242 & -0.0556 & -0.0099  \\ [1.7ex]
CCM : $3 \leq s \leq \frac{9}{2}$; cubic & -0.3149 & -0.0168 & -0.2574 & -0.0055 & -0.1137 & -0.0217 & -0.0507 & -0.0455 \\ [1.7ex] \hline \\ [-0.7ex]
SWT                                      & -0.3148 & -0.0165 & -0.2582 &         & -0.1150 &         & -0.0605 &       \\ [1.7ex] \hline\hline 
\end{tabular}
}
\end{center}
\label{tbl_low-E-param-ExtrapoFits}
\end{table}
For two of the fits we use the eight results with
$s=\{1,\,\frac{3}{2},\,2,\,\frac{5}{2},\,3,\,\frac{7}{2},\,4,\,\frac{9}{2}\}$,
for another two we use the six results with
$s=\{2,\,\frac{5}{2},\,3,\,\frac{7}{2},\,4,\,\frac{9}{2}\}$, while for
the last two we use the four results with
$s=\{3,\,\frac{7}{2},\,4,\,\frac{9}{2}\}$.  For each of these we fit
to the forms of Eqs.\
(\ref{honey-pure_E-fit_inversePower})--(\ref{honey-pure_X-fit_inversePower})
that are either quadratic or cubic in the parameter $1/s$.
For comparison we also show in Table \ref{tbl_low-E-param-ExtrapoFits}
results from SWT \cite{Zheng:1991_honey,Mattsson:1994_honey}.  Our
corresponding results for $E(s)/N$, $M(s)$, $\rho_{s}(s)$ and
$\chi(s)$ are also shown in Figs.\
\ref{E_multiSpins}--\ref{X_multiSpins}, respectively, where we again
compare with known results from SWT up to O($1/s^{n}$), denoted as
SWT($n$).

The SWT(2) results for the GS energy per spin, $E(s)/N$, shown in
Table \ref{tbl_low-E-param} and Fig.\ \ref{E_multiSpins}, are taken
from Zheng, Oitmaa and Hamer (ZOH) \cite{Zheng:1991_honey}.  ZOH also
give SWT(1) results for both the order parameter, $M(s)$, and the
zero-field transverse magnetic susceptibility, $\chi(s)$, and these
too are shown in both Table \ref{tbl_low-E-param-ExtrapoFits} and
Figs.\ \ref{M_multiSpins} and \ref{X_multiSpins}, respectively.  ZOH
do not cite SWT results for the spin stiffness, $\rho_{s}(s)$.  However,
Mattsson {\it et al}. \cite{Mattsson:1994_honey} cite the SWT(1)
result for the spin-wave velocity, $\hbar c(s) =
(3\sqrt{2}/2)J_{1}ds[1+0.20984/(2s)]$.  By making use of the
hydrodynamic relation of Eq.\ (\ref{eq_hbar_c}) and the ZOH SWT(1)
relation for $\chi(s)$, this readily yields the corresponding SWT(1)
relation, $\rho_{s}(s)=J_{1}d^{2}s^{2}(\frac{3}{4}-0.1150/s)$, which we
have shown in Table \ref{tbl_low-E-param-ExtrapoFits} and Fig.\
\ref{sStiff_multiSpins}.

We see from Table \ref{tbl_low-E-param-ExtrapoFits} that our
calculated coefficients for the GS energy, $\epsilon_{1}$ and
$\epsilon_{2}$, are in remarkable agreement with the corresponding
SWT(2) values.  Figure \ref{E_multiSpins}(b) shows very clearly how
simple $O(1/s^{n})$ fits with $n=2$ or $3$ to the CCM data points for
$E(s)/(NJ_{1}s^{2})$ with $s=\{3,\,\frac{7}{2},\,4,\,\frac{9}{2}\}$,
agree extremely well with both the corresponding SWT(2) result and the
unfitted CCM data points with $1 \leq s \leq \frac{5}{2}$.  Even the
extreme quantum case $s=\frac{1}{2}$ is rather well described by these
simple high-spin forms.  Table \ref{tbl_low-E-param-ExtrapoFits} also
shows that the leading-order coefficient for the magnetic order
parameter, $\mu_{1}$, as extracted from our extrapolated CCM results
for $M(s)$, is in very good agreement with that from SWT, and that the
next-to-leading coefficient, $\mu_{2}$, is small.  Figure
\ref{M_multiSpins}(b) shows too that simple $O(1/s^{n})$ fits with
$n=1,\,2$ or $3$ to the CCM data points for $M(s)/s$ with
$s=\{3,\,\frac{7}{2},\,4,\,\frac{9}{2}\}$, once again agree well with
both the corresponding SWT(1) result and the unfitted CCM data points
with $1 \leq s \leq \frac{5}{2}$.  However, by contrast with the GS
energy result, all of the fits lead to values for the
spin-$\frac{1}{2}$ case of rather limited accuracy.  Perhaps somewhat
counter-intuitively, the spin-$\frac{1}{2}$ model is actually {\it
  more} ordered than the (relatively low-order) high-spin expansions
would predict.

Turning to the spin stiffness, $\rho_{s}$, we see from Table
\ref{tbl_low-E-param-ExtrapoFits} that the leading quantum correction
coefficient, $\rho_{1}$, agrees extremely well with the SWT(1) result.
It is also clear that the second-order quantum correction coefficient,
$\rho_{2}$, is small and negative.  Our best estimates are obtained
from the fits to the higher-spin values, all of which are consistent
with a value $\rho_{2} \approx -0.025 \pm 0.004$.  Figure
\ref{sStiff_multiSpins}(b) shows that the quadratic and cubic fits to
the data points with $s=\{3,\,\frac{7}{2},\,4,\,\frac{9}{2}\}$ give
very good agreement with the unfitted CCM data points with
$s=\{1,\,\frac{3}{2},\,2,\,\frac{5}{2}\}$, with only the
$s=\frac{1}{2}$ point not fitted well by the high-spin expansion.

Finally, the results shown in Table \ref{tbl_low-E-param-ExtrapoFits}
lead to the observation that the low-energy parameter with the
greatest uncertainty associated with its high-spin expansion is the
transverse magnetic susceptibility, $\chi$.  While all of the values
obtained for the leading-order quantum correction coefficient,
$\chi_{1}$, are in reasonable agreement with the corresponding SWT(1)
result, the spread in the values is greater than for any of the other
low-energy parameters.  Similarly, while we can conclude that the
magnitude of the second-order coefficient $\chi_{2}$ is probably
smaller than (or, at most, comparable to) that of $\chi_{1}$, our CCM
fits do not allow us to predict the sign of $\chi_{2}$ with any real
degree of certainty.  It is interesting to note that this is exactly
what was also observed in a corresponding set of fits of the
low-energy parameters of the triangular-lattice HAF
\cite{Gotze:2016_triang} to their high-spin asymptotic expansions.
Figure \ref{X_multiSpins}(b) shows that the quadratic and cubic fits
to the CCM data points for $\chi$ with
$s=\{3,\,\frac{7}{2},\,4,\,\frac{9}{2}\}$ now give good agreement only
with the unfitted data points with $s=\{2,\,\frac{5}{2}\}$, with a
discrepancy already opening up at the value $s=\frac{3}{2}$.

\section{SUMMARY}
\label{summary_sec}
In two dimensions the honeycomb lattice has the smallest coordination
number ($z=3$), and the effects of quantum fluctuations are hence the
greatest.  Thus, the HAF on the honeycomb lattice occupies a special
place in the field of theoretical quantum magnetism.  Nevertheless,
there exist very few studies of this model that examine within a
coherent and unified framework the role of the spin quantum number $s$ on its
low-energy properties.  Furthermore, there also exists a rather large
number of experimental realizations of quasi-2D honeycomb-lattice
systems with AFM interactions and with various values of $s$.

For example, the magnetic compounds InCu$_{2/3}$V$_{1/3}$O$_{3}$
\cite{Kataev:2005_honey}, Na$_{3}$Cu$_{2}$SbO$_{6}$
\cite{Miura:2006_honey}, $\beta$-Cu$_{2}$V$_{2}$O$_{7}$
\cite{Tsirlin:2010_honey} and Cu$_{5}$SbO$_{6}$
\cite{Climent:2012_honey} all contain $s=\frac{1}{2}$ Cu$^{2+}$ ions
situated on the sites of weakly-coupled honeycomb-lattice layers.
Similarly, the iridate family of compounds A$_{2}$IrO$_{3}$ (A $=$
Na, Li)
\cite{Singh:2010_honey,Liu:2011_honey,Singh:2012_honey,Choi:2012_honey}
are believed to be magnetically ordered Mott insulators in which
the Ir$^{4+}$ ions, that are also arrayed on weakly-coupled
honeycomb-lattice layers, form effective $s=\frac{1}{2}$ moments.  The
families of compounds BaM$_{2}$(XO$_{4}$)$_{2}$ (M $=$ Co, Ni; X $=$
P, As) \cite{Regnault:1990_honey} and Cu$_{3}$M$_{2}$SbO$_{6}$ (M $=$
Co, Ni) \cite{Roudebush:2013_honey} also comprise similar
honeycomb-lattice materials.  The magnetic M$^{2+}$ ions in both
families again occupy the sites of a honeycomb lattice in layers that
are weakly coupled.  For both families, when M $=$ N$_{i}$, the
Ni$^{2+}$ ions appear to take the high-spin value (viz., $s=1$) in
both cases.  By contrast, when M $=$ Co, whereas the Co$^{2+}$ ions
appear to take the low-spin value (viz., $s=\frac{1}{2}$) in the
former family BaCo$_{2}$(XO$_{4}$)$_{2}$, and the high-spin value
(viz., $s=\frac{3}{2}$) in the latter compound
Cu$_{3}$Co$_{2}$SbO$_{6}$.  Another example of a spin-$\frac{3}{2}$
honeycomb-lattice AFM material is the layered compound
Bi$_{3}$Mn$_{4}$O$_{12}$(NO$_{3}$)
\cite{Smirnova:2009:honey_spin_3half,Okubo:2010:honey_spin_3half} in
which the Mn$^{4+}$ ions occupy the sites of the
honeycomb-lattice layers and take the value $s=\frac{3}{2}$.

We have presented here large-scale, high-order CCM calculations for
the complete set of low-energy GS parameters (viz., the energy per
spin $E/N$, sublattice magnetization $M$, spin stiffness $\rho_{s}$
and transverse zero-field magnetic susceptibility $\chi$) for the honeycomb
lattice HAF, for values of the spin quantum number $s$ in the range
$\frac{1}{2} \leq s \leq \frac{9}{2}$.  The {\it only} approximation
made in our CCM calculations has been the truncation of the resulting
coupled sets of equations for the GS correlation coefficients that
completely determine all GS properties, within a systematic
SUB$n$--$n$ hierarchy of approximations that becomes asymptotically
exact as the truncation parameter $n \rightarrow \infty$.  We have
performed calculations for arbitrary spin quantum number $s$ to high
orders in $n$ (viz., for $n \leq 10$ for calculations of the
parameters $E/N$ and $M$, and $n \leq 8$ for calculations of the
parameters $\rho_{s}$ and $\chi$), and have extrapolated these to the
exact $n \rightarrow \infty$ limit in each case, using
well-tested extrapolation schemes, thereby obtaining results of proven
high accuracy.  We have used the CCM results with the larger values of
$s$ to derive high-spin asymptotic series for the low-energy
parameters, and have compared these with corresponding results from
SWT, where available.  We have also shown explicitly how the extreme
quantum cases $s=\frac{1}{2}$ and $s=1$ can deviate appreciably from
the behaviour predicted by such large-$s$ expansions.

\section*{ACKNOWLEDGMENTS}
We thank the University of Minnesota Supercomputing Institute for the
grant of the supercomputing facilities on which this work was
performed.  One of us (RFB) gratefully acknowledges the Leverhulme
Trust for the award of an Emeritus Fellowship (EM-2015-07).

\section*{References}



\bibliographystyle{elsarticle-num} 
\bibliography{bib_general}





\end{document}